\documentclass[lettersize,journal]{IEEEtran}
\usepackage{amsmath,amsfonts}
\usepackage{algorithmic}
\usepackage{algorithm}
\usepackage{array}
\usepackage[caption=false,font=normalsize,labelfont=sf,textfont=sf]{subfig}
\usepackage{textcomp}
\usepackage{stfloats}
\usepackage{url}
\usepackage{verbatim}
\usepackage{graphicx}
\usepackage{cite}
\usepackage{bm}
\usepackage{amssymb}
\begin{document}

\title{Active RIS-Assisted mmWave Indoor Signal Enhancement Based on Transparent RIS}

\author{Hao Feng, Yuping Zhao
\thanks{(Corresponding author: Yuping Zhao.)}
\thanks{Hao Feng is with the Peking University Shenzhen Graduate School, Peking University, Shenzhen 518066, China, with the Peng Cheng Laboratory, Shenzhen 518066, China, and also with the School of Electronics, Peking University, Beijing 100871, China (e-mail: hfeng@pku.edu.cn).}
\thanks{Yuping Zhao is with the School of Electronics, Peking University, Beijing 100871, China (e-mail: yuping.zhao@pku.edu.cn).}}

\markboth{Journal of \LaTeX\ Class Files,~Vol.~14, No.~8, August~2021}%
{Shell \MakeLowercase{\textit{et al.}}: A Sample Article Using IEEEtran.cls for IEEE Journals}


\maketitle
\begin{abstract}
Due to the serious path loss of millimeter-wave (mmWave), the signal sent by the base station is seriously attenuated when it reaches the indoors. Recent studies have proposed a glass-based metasurface that can enhance mmWave indoor signals. The transparent reconfigurable intelligent surface (RIS) focuses on the mmWave signal to a specific location indoors. In this paper, a novel RIS-assisted mmWave indoor enhancement scheme is proposed, in which a transparent RIS is deployed on the glass to enhance mmWave indoor signals, and three assisted transmission scenarios, namely passive RIS (PRIS), active RIS (ARIS), and a novel hybrid RIS (HRIS) are proposed. This paper aims to maximize the signal-to-noise ratio (SNR) of the received signal for the three assisted transmission scenarios. The closed-form solution to the maximum SNR is presented in the PRIS and the ARIS-assisted transmission scenarios. Meanwhile, the closed-form solution to the maximum SNR for the HRIS-assisted transmission scenario is presented for given active unit cells. In addition, the performance of the proposed scheme is analyzed under three assisted transmission scenarios. The results indicate that under a specific RIS power budget, the ARIS-assisted transmission scenario achieves the highest data rate and energy efficiency. Also, it requires very few unit cells, thus dramatically reducing the size of the metasurface.
\end{abstract}


\section{Introduction}
\IEEEPARstart{W}{ith} the large-scale commercialization of the fifth-generation (5G) communication technology, the 5G platform supports a broad range of high data-rate applications~\cite{ref01}, which have a high demand for network capacity~\cite{ref02,ref03}. Meanwhile, with the emergence of more and more services requiring high data rates and the continuous increase in the number of mobile-connected devices~\cite{ref04}, the demand for data rates has seen a great increase~\cite{ref05}. To support new mobile application scenarios, new wireless communication standards for higher frequencies should be formulated, such as the millimeter wave (mmWave)~\cite{ref06,ref07}. The mmWave band can provide more bandwidth and serve high data-rate applications. However, the mmWave band has severe path loss~\cite{ref08}. The obstacles, such as concrete, can hinder mmWave signals to propagate from the outdoor base station (BS) to the indoor user equipment (UE). The traditional approach to overcome this propagation limitation is to use relays between the BS and the UE~\cite{ref09}. \cite{ref10} proposed using a decode-forward relay for indoor signal enhancement of mmWave and conducted hardware experiments.

Another approach for solving this problem is to use a reconfigurable intelligent surface (RIS)~\cite{ref11,ref12,ref13,ref14}. RIS is composed of many small, low-cost unit cells, each of which can change the phase shift of the incident signal to reflect or transmit the incident signal, thereby changing the radio wave propagation direction~\cite{ref15,ref16,ref17,ref18}, enhancing user signal quality~\cite{ref19,ref20} and improving communication system performance~\cite{ref21,ref22,ref23}. Recent research investigated transparent RIS (transmission RIS). NTT DOCOMO~\cite{ref24,ref25} is the first company to study transparent RIS and designed a glass-based metasurface that can dynamically control scattering characteristics and realize mmWave focusing while achieving a large area and high transparency.

However, when the signal transmitted by the BS is concentrated on a specific position indoors through the transparent RIS, the signal connection needs to transmit the signals to the UE. Reflection RIS can be used for indoor signal connections. In most of the existing studies, the reflection RIS unit cell is passive, i.e., passive RIS (PRIS). Because of the multiplicative effect of channel fading~\cite{ref26,ref27}, the reflected signal must pass through a cascaded channel, i.e., the path loss of the BS-transmission RIS-PRIS-UE link is the product rather than the sum. So, PRIS requires a large number of unit cells to increase the channel gain. Recent studies have proposed an active reflection RIS (ARIS) to alleviate the multiplicative effect of channel fading. ARIS can amplify the reflected signal to handle the problem~\cite{ref28}, but it uses many power amplifiers and increases the hardware cost. ARIS is generally composed of several active reflective unit cells equipped with negative resistance elements such as tunnel diodes and negative impedance converters~\cite{ref29,ref30,ref31}, which can amplify the reflected power of incident signals~\cite{ref32,ref33}. In~\cite{ref26}, the authors proposed to use ARIS to assist wireless communication, and they compared the theoretical performance of ARIS and PRIS. In~\cite{ref27}, the authors implemented a unit cell of ARIS and tested the unit cell performance. In~\cite{ref28}, the authors studied the deployment of ARIS and compared the performance of ARIS and PRIS under different deployment strategies. To trade off cost and channel gain, this paper proposes a hybrid reflection RIS (HRIS), which includes both passive and active unit cells. HRIS can use fewer active unit cells to achieve higher channel gain without generating more hardware costs. 
\begin{figure}[!t]
\centering
\includegraphics[width=3in]{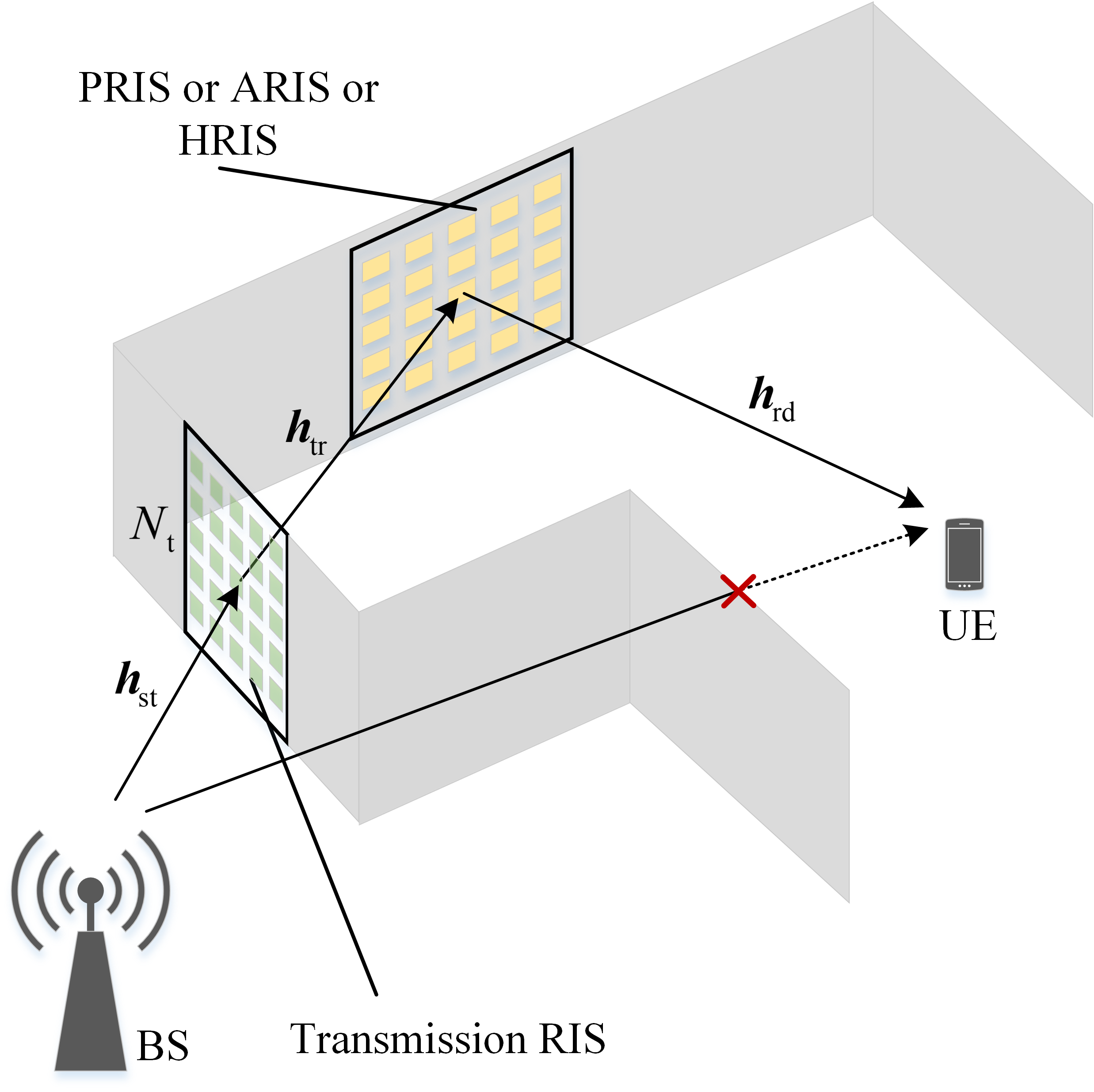}
\caption{Schematic diagram of the system model.}
\label{fig01}
\end{figure}

Meanwhile, a novel RIS-assisted mmWave indoor enhancement scheme is proposed, in which a transparent RIS is deployed on the glass to enhance the mmWave indoor signal, focus the mmWave signal emitted by an outdoor BS to a specific indoor location, and uses PRIS, ARIS, and HRIS deployed indoors to transmit the signal to the UE. First, this paper maximizes the UE's signal-to-noise ratio (SNR) under the three assisted transmission scenarios to maximize the data rate. The closed-form solution of the maximum SNR is presented in the PRIS and ARIS-assisted transmission scenarios. Then, the maximum SNR closed-form solution is presented for given active unit cells in the HRIS-assisted transmission scenario. The results show that the achievable data rate in the ARIS-assisted transmission scenario is the largest under a specific RIS power budget and the ARIS-assisted transmission scenario achieves the highest energy efficiency.

The main contributions of this paper are summarized as follows:
\begin{itemize}
\item A novel RIS-assisted mmWave indoor enhancement scheme is proposed. In this scheme, a transparent RIS is deployed on the glass to enhance mmWave indoor signals, and three assisted transmission scenarios of PRIS, ARIS, and HRIS are used indoors. Also, HRIS is first proposed here.
\item For the proposed scheme, the problem of maximizing the SNR of the received signal is analyzed to maximize the achievable rate. The closed form of the maximum SNR is presented for PRIS and ARIS-assisted transmission scenarios. In the HRIS-assisted transmission scenario, the closed-form solution of the maximum SNR is presented for given active unit cells.
\item The performance of the proposed scheme is analyzed under three assisted transmission scenarios. The results show that the achievable data rate in the ARIS-assisted transmission scenario is the largest under a specific RIS power budget, and the number of unit cells required is the smallest. Under the same RIS power budget, the achievable data rate of the HRIS-assisted is higher than that of the PRIS-assisted transmission scenario.
\item The energy efficiency of using the three assisted transmission scenarios is compared. The results show that the ARIS-assisted transmission scenario achieves the highest energy efficiency, and the energy efficiency of the HRIS-assisted is higher than that of the PRIS-assisted transmission scenario.
\end{itemize}

The rest of this paper is organized as follows. The system model of the proposed RIS-assisted mmWave indoor enhancement scheme is introduced in Section II, and the indoor signal connection adopts PRIS, ARIS, and HRIS, respectively. In Section III, the SNR maximization problem is formulated under the three assisted transmission scenarios. Section IV analyzes the performance of the proposed scheme under the three assisted transmission scenarios. In Section V, numerical simulations are conducted to evaluate the performance of the proposed scheme in different communication scenarios. Finally, the main achievements and conclusions are presented in Section VI.

The main symbols used in this paper are as follows: the lowercase letter $x$ represents a scalar; the bold letter represents a vector or matrix; $|x|$ represents the absolute value of $x$; $|\bm{x}|$ represents the absolute value of the vector element of $\bm{x}$, and $\left\| \bm{x} \right\|$ represents the norm of the vector $\bm{x}$; ${\mathcal C}{\mathcal N}\left( {\mu ,{\delta ^2}} \right)$ represents a complex Gaussian distribution with a mean of $\mu $ and a variance of ${\delta ^2}$; $\bm{x}^T$ represents the transpose of $\bm{x}$.

\section{System Model}
Fig.~\ref{fig01} shows the diagram of the system model. The UE is indoor and far away from the BS. Due to the severe path loss of mmWave, this paper considers deploying ${N_{\text{t}}}$ passive transmission unit cells on the window to enhance the indoor mmWave signal. Each unit cell can induce an independent phase shift to the transmitted signal, thereby cooperatively changing the effective channel from the BS to UE. To increase the coverage of the signal and transmit the signals to the UE, signal connections are needed. This paper considers using ARIS, PRIS, and HRIS-assisted transmission scenarios. In Fig.~\ref{fig01}, the system consists of a UE, a BS, a transmission RIS, and three types of RIS, wherein the UE and the BS are configured with an antenna. Due to obstacles such as concrete, the direct path signal is feeble, so this paper does not consider the direct path here.

\subsection{Channel model}
Assume that the number of unit cells of PRIS, ARIS, and HRIS are ${N_{\text{p}}}$, ${N_{\text{a}}}$ and ${N_{\text{h}}}$, respectively. In the HRIS, ${N_{\text{h}}} = {N_{{\text{ha}}}} + {N_{{\text{hp}}}}$, where ${N_{{\text{hp}}}}$ and ${N_{{\text{ha}}}}$ are passive and active unit cells, respectively. ${\bm{h}_{{\text{st}}}} \in {{\mathbb C}^{{N_{\text{t}}}}}$ denotes the channel vectors from BS to transmission RIS. $\bm{h}_{{\text{tr}}}^k \in {{\mathbb C}^{{N_k} \times {N_{\text{t}}}}}$, $k \in {\mathcal K} \triangleq \left \{ \text{p,a,hp,ha} \right \}$, denotes the channel matrixes from transmission RIS to PRIS, ARIS, passive unit cells in HRIS, and active unit cells in HRIS, respectively. The deterministic flat-fading channels are considered here. $\bm{h}_{{\text{rd}}}^k \in {{\mathbb C}^{{N_k}}}$, $k \in {\mathcal K}$ denotes the channel vectors from PRIS, ARIS, passive unit cells in HRIS, and active unit cells in HRIS to UE, respectively. ${\bm{\Phi} _{\text{t}}} = {\text{diag}}\left\{ {\phi _1^{\text{t}},\phi _2^{\text{t}}, \cdots , \phi _{{N_{\text{t}}}}^{\text{t}}} \right\} \in {{\mathbb C}^{{N_{\text{t}}} \times {N_{\text{t}}}}}$ denotes the transmission coefficient matrix of the transmission RIS, where $\phi _i^{\text{t}}{ = }{\alpha _{\text{t}}}{e^{j\theta _i^{\text{t}}}}$, ${\alpha _{\text{t}}} \in \left( {0,1} \right]$ is the magnitude. $\theta _i^{\text{t}} \in \left( 0, 2 \pi \right ] $ is the phase shift variable, $i \in \left\{ {1, \cdots ,{N_{\text{t}}}} \right\} \triangleq {{\mathcal N}_{\text{t}}}$. ${\bm{\Phi} _k} = {\text{diag}}\left\{ {\phi _1^k,\phi _2^k, \cdots, \phi _{{N_k}}^k} \right\} \in {{\mathbb C}^{{N_k} \times {N_k}}}$, $k \in {\mathcal K}$ is the reflection coefficient matrix of PRIS, ARIS, passive unit cells in HRIS, and active unit cells in HRIS, respectively, where $\phi _j^k{ = }{\alpha _k}{e^{j\theta _j^k}}$, ${\alpha _k} \in \left( {0,1} \right]$ is the magnitude, and $\theta _j^k \in \left( 0, 2 \pi \right ] $ is the phase shift variable, $j \in \left\{ {1, \cdots ,{N_k}} \right\} \triangleq {{\mathcal N}_k}$. Here, this paper assumes that the coefficient magnitude of the transmission RIS, PRIS, ARIS, passive unit cells in HRIS, and active unit cells in HRIS is 1, i.e., ${\alpha _{\text{t}}} = {\alpha _k} = 1$, $k \in {\mathcal K}$.

\subsection{Signal model}
Let $s$ be the information symbol sent by BS and $s$ be a complex Gaussian random variable with zero mean and unit variance. The transmitted signals of the BS can be expressed as
\begin{equation}
\label{eq-1}
x = \sqrt p s,
\end{equation}
where $p$ is the transmit power of the BS. The received signal at the UE using PRIS for assisted transmission scenario can be expressed as
\begin{equation}
\label{eq-2}
{y_{\text{p}}} = \left( {\bm{h}{{_{{\text{rd}}}^{\text{p}}}^T}{\bm{\Phi} _{\text{p}}}\bm{h}_{{\text{tr}}}^{\text{p}}{\bm{\Phi} _{\text{t}}}{\bm{h}_{{\text{st}}}}} \right)x + n,
\end{equation}
where $n \sim {\mathcal C}{\mathcal N}\left( {0,{\delta ^2}} \right)$ represents additive white Gaussian noise with an average power of ${\delta ^2}$. The SNR at the UE using PRIS for assisted transmission scenario can be expressed as
\begin{equation}
\label{eq-3}
{\gamma _{\text{P}}} = \frac{{p{{\left| {\bm{h}{{_{{\text{rd}}}^{\text{p}}}^T}{\bm{\Phi} _{\text{p}}}\bm{h}_{{\text{tr}}}^{\text{p}}{\bm{\Phi} _{\text{t}}}{\bm{h}_{{\text{st}}}}} \right|}^2}}}{{{\delta ^2}}}.
\end{equation}

In ARIS-assisted transmission scenario, ARIS receives the signal, amplifies the received signal by $\beta $ times, and transmits it in a full-duplex form. Then, the received signal at the UE can be expressed as
\begin{equation}
\label{eq-4}
{y_{\text{a}}} = {\bm{h}_{\text{rd}}^\text{a}}^T {\bm{\Phi} _{\text{a}}}\beta \left( {\bm{h}_{{\text{tr}}}^{\text{a}}{\bm{\Phi} _{\text{t}}}{\bm{h}_{{\text{st}}}}x + {\bm{n}_1}} \right) + {n_2},
\end{equation}
where ${\bm{n}_1} \sim {\mathcal C}{\mathcal N}\left( {\bm{0},{\delta ^2}{\bm{I}_{{N_{\text{a}}}}}} \right)$, ${n_2} \sim {\mathcal C}{\mathcal N}\left( {0,{\delta ^2}} \right)$ represent noise, and the average power is $\delta ^2$. The SNR at the UE using ARIS for assisted transmission scenario can be expressed as
\begin{equation}
\label{eq-5}
{\gamma _{\text{A}}} = \frac{{p{{\left| {\bm{h}{{_{{\text{rd}}}^{\text{a}}}^T}\beta {\bm{\Phi} _{\text{a}}}\bm{h}_{{\text{tr}}}^{\text{a}}{\bm{\Phi} _{\text{t}}}{\bm{h}_{{\text{st}}}}} \right|}^2}}}{{\left( {{{\left\| {\bm{h}{{_{{\text{rd}}}^{\text{a}}}^T}\beta {\bm{\Phi} _{\text{a}}}} \right\|}^2} + 1} \right){\delta ^2}}}.
\end{equation}

In HRIS-assisted transmission scenario, the received signal at the UE can be expressed as
\begin{equation}
\label{eq-6}
{y_{\text{h}}} = {\bm{h}_{\text{rd}}^\text{ha}}^T \beta {\bm{\Phi} _{{\text{ha}}}}\left( {\bm{h}_{{\text{tr}}}^{{\text{ha}}}{\bm{\Phi} _{\text{t}}}{\bm{h}_{{\text{st}}}}x + {\bm{n}_3}} \right) + {\bm{h}_{{\text{rd}}}^{{\text{hp}}}}^T{\bm{\Phi} _{{\text{hp}}}}\bm{h}_{{\text{tr}}}^{{\text{hp}}}{\bm{\Phi} _{\text{t}}}{\bm{h}_{{\text{st}}}}x + {n_4},
\end{equation}
where ${\bm{n}_3} \sim {\mathcal C}{\mathcal N}\left( {\bm{0},{\delta ^2}{\bm{I}_{{N_{{\text{ha}}}}}}} \right)$, ${n_4} \sim {\mathcal C}{\mathcal N}\left( {0,{\delta ^2}} \right)$represents noise, and the average power is ${\delta ^2}$. The SNR of using HRIS for assisted transmission scenario can be expressed as
\begin{equation}
\label{eq-7}
{\gamma _{\text{H}}} = \frac{{p{{\left| {\bm{h}{{_{{\text{rd}}}^{{\text{hp}}}}^T}{\bm{\Phi} _{{\text{hp}}}}\bm{h}_{{\text{tr}}}^{{\text{hp}}}{\bm{\Phi} _{\text{t}}}{\bm{h}_{{\text{st}}}} + \bm{h}{{_{{\text{rd}}}^{{\text{ha}}}}^T}\beta {\bm{\Phi} _{{\text{ha}}}}\bm{h}_{{\text{tr}}}^{{\text{ha}}}{\bm{\Phi} _{\text{t}}}{\bm{h}_{{\text{st}}}}} \right|}^2}}}{{\left( {{{\left\| {\bm{h}{{_{{\text{rd}}}^{{\text{ha}}}}^T}\beta {\bm{\Phi} _{{\text{ha}}}}} \right\|}^2} + 1} \right){\delta ^2}}}.
\end{equation}

\subsection{Power consumption at the RIS}
This subsection discusses the power consumption of PRIS, ARIS, and HRIS. The power consumption of PRIS mainly comes from the switching and control circuits of unit cells~\cite{ref37,ref38}, and the power dissipation of PRIS can be expressed as
\begin{equation}
\label{eq-8}
{P_{{\text{PRIS}}}} = {N_{\text{p}}}{P_{\text{e}}},
\end{equation}
where $ P_{\text{e}}$ is the power dissipation of each unit cell due to the circuit consumption required for adaptive phase shifting. For the power consumption of ARIS, the typical power amplification model of ARIS~\cite{ref39} can be represented as
\begin{equation}
\label{eq-9}
{P_{{\text{ARIS}}}} = {\text{OPI}} + {\text{OPD}},
\end{equation}
where OPI includes the power consumption of the switch and control circuit and the power consumption of DC bias voltage. OPI can be expressed as
\begin{equation}
\label{eq-10}
{\text{OPI}} = {N_{\text{a}}}\left( {{P_{\text{e}}} + {P_{{\text{DC}}}}} \right),
\end{equation}
where ${P_{{\text{DC}}}}$ is the power of the DC bias voltage. OPD can be represented as a linear model
\begin{equation}
\label{eq-11}
{\text{OPD}} = \frac{{{P_{{\text{out,a}}}}}}{v},
\end{equation}
where ${P_{{\text{out,a}}}}$ is the amplification power of ARIS, and $v$ is the efficiency of the power amplifier. The relationship between the output power and incident signal power ${P_{{\text{in,a}}}}$ can be expressed as
\begin{equation}
\label{eq-12}
\begin{aligned}
P_{{\text{out,a}}} & = {\beta ^2}{P_{{\text{in,a}}}}  \\
& = {\beta ^2}\left( {p{{\left\| {{\bm{\Phi} _{\text{a}}}\bm{h}_{{\text{tr}}}^{\text{a}}{\bm{\Phi} _{\text{t}}}{\bm{h}_{{\text{st}}}}} \right\|}^2}{\text{ + }}{\delta ^2}{{\left\| {{\bm{\Phi} _{\text{a}}}{\bm{I}_{{N_{\text{a}}}}}} \right\|}^2}} \right) .
\end{aligned}
\end{equation}

This paper assumes that the amplification factor of all unit cells is $\beta $, and the reflection amplifier works in a linear region, i.e., the output power increases linearly with the input power. The transmitted signal from the BS undergoes double attenuation in the BS-transmission RIS and transmission RIS-ARIS channels before reaching each unit cell of the ARIS. Thus, the incident signal has a weak power, and the reflection amplifier can obtain a higher amplification gain at a lower output power. The total power of ARIS can be expressed as
\begin{equation}
\label{eq-13}
{P_{{\text{ARIS}}}} = {N_{\text{a}}}\left( {{P_{\text{e}}} + {P_{{\text{DC}}}}} \right) + \frac{{{P_{{\text{out,a}}}}}}{v}.
\end{equation}

For a given power budget ${P_{{\text{ARIS}}}}$, after allocating hardware power for the unit cells, ARIS allocates the remaining power for active load amplification of the incident signal. For the maximum amplification factor ${\beta _{\max }}$ that an active load can provide, there are the following amplification constraints
\begin{equation}
\label{eq-14}
\beta  \leqslant {\beta _{\max }}.
\end{equation}

Meanwhile, the amplification power is also limited due to the power budget. When designing the phase shift matrices for transmission RIS and ARIS $\left\{ {{\bm{\Phi} _{\text{t}}},{\bm{\Phi} _{\text{a}}}} \right\}$, the amplification power limit can be expressed as
\begin{equation}
\label{eq-15}
{\beta ^2}\left( {p{{\left\| {{\bm{\Phi} _{\text{a}}}\bm{h}_{{\text{tr}}}^{\text{a}}{\bm{\Phi} _{\text{t}}}{\bm{h}_{{\text{st}}}}} \right\|}^2}{\text{ + }}{\delta ^2}{{\left\| {{\bm{\Phi} _{\text{a}}}{\bm{I}_{{N_{\text{a}}}}}} \right\|}^2}} \right) \leqslant P_{\text{a}},
\end{equation}
where the amplification power budget of ARIS is defined as
\begin{equation}
\label{eq-16}
{P_{\text{a}}} \triangleq \left( {{P_{{\text{ARIS}}}} - {N_{\text{a}}}\left( {{P_{\text{e}}} + {P_{{\text{DC}}}}} \right)} \right)v.
\end{equation}

Similar to ARIS, the total power of HRIS can be expressed as
\begin{equation}
\label{eq-17}
\begin{aligned}
&{P_{{\text{HRIS}}}} = \\
&{N_{\text{h}}}{P_{\text{e}}} + {N_{{\text{ha}}}}{P_{{\text{DC}}}} + \frac{{{\beta ^2}\left( {p{{\left\| {{\bm{\Phi} _{{\text{ha}}}}\bm{h}_{{\text{tr}}}^{{\text{ha}}}{\bm{\Phi} _{\text{t}}}{\bm{h}_{{\text{st}}}}} \right\|}^2}{\text{ + }}{\delta ^2}{{\left\| {{\bm{\Phi} _{{\text{ha}}}}{\bm{I}_{{N_{{\text{ha}}}}}}} \right\|}^2}} \right)}}{v},
\end{aligned}
\end{equation}
where the amplification power budget of HRIS is defined as
\begin{equation}
\label{eq-18}
{P_{\text{h}}} \triangleq \left( {{P_{{\text{HRIS}}}} - {N_{\text{h}}}{P_{\text{e}}} - {N_{{\text{ha}}}}{P_{{\text{DC}}}}} \right)v.
\end{equation}

\section{Theoretical Study of RIS-assisted Transmission Scenario}
In this section, the problem of maximizing the achievable data rate for UE is analyzed under RIS-assisted transmission scenario.

\subsection{PRIS-assisted transmission scenario}
Our goal is to maximize the downlink achievable data rate of the PRIS-assisted transmission scenario by optimizing the transmission RIS and PRIS phase shift matrices $\left\{ {{\bf{\Phi} _{\text{t}}},{\bm{\Phi} _{\text{p}}}} \right\}$. The corresponding formula is as follows
\begin{align}
(\text{\bf{P1}}): \mathop {\max }\limits_{{\bm{\Phi} _{\text{t}}},{\bm{\Phi} _{\text{p}}}} &{\log _2}\left( {1 + \frac{{p{{\left| {\bm{h}{{_{{\text{rd}}}^{\text{p}}}^T}{\bm{\Phi} _{\text{p}}}\bm{h}_{{\text{tr}}}^{\text{p}}{\bm{\Phi} _{\text{t}}}{\bm{h}_{{\text{st}}}}} \right|}^2}}}{{{\delta ^2}}}} \right),\label{eq-19}\\
\text{s.t.} &\left| {{{\left[ {{\bm{\Phi} _{\text{t}}}} \right]}_i}} \right| = 1,\forall i \in {{\mathcal N}_{\text{t}}},\tag{\ref{eq-19}{a}} \label{eq-19a}\\
&\left| {{{\left[ {{\bm{\Phi} _{\text{p}}}} \right]}_j}} \right| = 1,\forall j \in {{\mathcal N}_{\text{p}}}.\tag{\ref{eq-19}{b}} \label{eq-19b}
\end{align}

(19) indicates that maximizing the downlink achievable data rate of the PRIS-assisted transmission scenario is equivalent to maximizing the SNR of the UE, and (P1) can be transformed into the following expression
\begin{align}
(\text{\bf{P2}}): \mathop {\max }\limits_{{\bm{\Phi} _{\text{t}}},{\bm{\Phi} _{\text{p}}}} &\frac{{p{{\left| {\bm{h}{{_{{\text{rd}}}^{\text{p}}}^T}{\bm{\Phi} _{\text{p}}}\bm{h}_{{\text{tr}}}^{\text{p}}{\bm{\Phi} _{\text{t}}}{\bm{h}_{{\text{st}}}}} \right|}^2}}}{{{\delta ^2}}},\label{eq-20}\\
\text{s.t.} & (\text{19a})-(\text{19b}).\notag
\end{align}

Then, the design of passive beamforming $\left\{ {{\bm{\Phi} _{\text{t}}},{\bm{\Phi} _{\text{p}}}} \right\}$ is discussed. To design the transmission and reflection coefficient matrices for optimal performance, this paper assumes that the channel state information (CSI) of the channel vector from the BS to the transmission RIS ${\bm{h}_{{\text{st}}}}$ and from the PRIS to the UE $\bm{h}_{{\text{rd}}}^{\text{p}}$ are known by the BS and the UE through channel estimation. However, since both ends of the transmission RIS and PRIS channels are passively reflected, pilot signals cannot be transmitted for channel estimation, so it is difficult to estimate the channel directly. However, since the transmission RIS and PRIS positions are fixed, \cite{ref34} proposes a geometric relationship-based method to characterize the line-of-sight (LOS) channel from the transmission RIS to the PRIS $\bm{h}_{{\text{tr}}}^{\text{p}}$, followed by a joint passive beamforming design.

According to~\cite{ref35} and~\cite{ref36}, the LOS channel from transmission RIS to PRIS can be assumed to be a rank-one matrix if the following expression holds
\begin{equation}
\label{eq-21}
{d_{{\text{tr}}}} \gg \frac{{\sqrt {{N_{\text{p}}}} {l^2}}}{\lambda },
\end{equation}
where ${d_{{\text{tr}}}}$ is the distance from the transmission RIS to the PRIS, $l$ is the distance between the adjacent unit cell, and $\lambda $ is the carrier wavelength. (21) usually holds because the distance between the transmission RIS and the PRIS is usually much larger than their size. According to the geometric relationship between transmission RIS and PRIS from~\cite{ref34}, the channel matrix $\bm{h}_{{\text{tr}}}^{\text{p}}$ can be decomposed into the product of two eigenvectors ${\bm{g}_{\text{p}}}$ and ${\bm{g}_{\text{t}}}$, namely
\begin{equation}
\label{eq-22}
\bm{h}_{{\text{tr}}}^{\text{p}} = q{\bm{g}_{\text{p}}}\bm{g}_{\text{t}}^T,
\end{equation}
where $q$ denotes the complex-valued path gain; ${\bm{g}_{\text{p}}} \in {{\mathbb C}^{{N_{\text{p}}}}}$ and ${\bm{g}_{\text{t}}} \in {{\mathbb C}^{{N_{\text{t}}}}}$ denote element-wise transmission and reflection vectors at transmission RIS and PRIS, respectively. Based on (22), the CSI from transmission RIS to PRIS can be calculated according to the geometric relationship between them.

Since the rank of $\bm{h}_{{\text{tr}}}^{\text{p}}$ is 1 under the LOS channel, the following approximate passive beamforming design can be obtained. Assuming that the channel from transmission RIS to PRIS is LOS and the distance from transmission RIS to PRIS is far enough, the transmission coefficient matrix of transmission RIS and the reflection coefficient matrix of PRIS can be configured as follows
\begin{equation}
\label{eq-23}
\phi _i^{\text{t}} = {\left( {\frac{{{{\left[ {{\bm{g}_{\text{t}}}} \right]}_i}{{\left[ {{\bm{h}_{{\text{st}}}}} \right]}_i}}}{{\left| {{{\left[ {{\bm{g}_{\text{t}}}} \right]}_i}} \right|\left| {{{\left[ {{\bm{h}_{{\text{st}}}}} \right]}_i}} \right|}}} \right)^*},{\text{ }}\;i \in {{\mathcal N}_{\text{t}}},
\end{equation}
\begin{equation}
\label{eq-24}
\phi _j^{\text{p}} = {\left( {\frac{{{{\left[ {\bm{h}_{{\text{rd}}}^{\text{p}}} \right]}_j}{{\left[ {{\bm{g}_{\text{p}}}} \right]}_j}}}{{\left| {{{\left[ {\bm{h}_{{\text{rd}}}^{\text{p}}} \right]}_j}} \right|\left| {{{\left[ {{\bm{g}_{\text{p}}}} \right]}_j}} \right|}}} \right)^*},{\text{ }}\;j \in {{\mathcal N}_{\text{p}}},
\end{equation}
where ${\left(  \cdot  \right)^{\text{*}}}$is the conjugate transpose.

The transmission RIS uses a passive beamformer shown in (23) to transmit the signals from the BS and then transmits them to a specific unit cell on the PRIS. PRIS uses a passive beamformer shown in (24) to further reflect the signals from the transmission RIS and transmit them to the UE.

According to the above results, the transmission coefficient matrix of transmission RIS and the reflection coefficient matrix of PRIS are substituted into (20). Then, the UE SNR of maximizing the PRIS-assisted transmission scenario can be expressed as
\begin{equation}
\label{eq-25}
{\gamma _{\text{P}}} = \frac{{p{{\left( {\sum\limits_{m = 1}^{{N_{\text{t}}}} {\sum\limits_{n = 1}^{{N_{\text{p}}}} {\left| {\bm{h}_{{\text{rd}}n}^{\text{p}}} \right|} \left| {\bm{h}_{{\text{tr}}nm}^{\text{p}}} \right|\left| {{\bm{h}_{{\text{st}}m}}} \right|} } \right)}^2}}}{{{\delta ^2}}}.
\end{equation}

Because the SNR is only related to the amplitude of the channel vector but not related to the phase. For simplification, this paper introduces the following representations $\left| {{h_{{\text{st}}}}} \right|{ = }\sqrt {{\beta _{{\text{st}}}}} $, $\left| {{h_{{\text{tr}}}}} \right|{ = }\sqrt {{\beta _{{\text{tr}}}}} $, $\left| {{h_{{\text{rd}}}}} \right|{ = }\sqrt {{\beta _{{\text{rd}}}}} $, and (25) can be rewritten as
\begin{equation}
\label{eq-26}
{\gamma _{\text{P}}} = \frac{{pN_{\text{t}}^2N_{\text{p}}^2{\beta _{{\text{rd}}}}{\beta _{{\text{tr}}}}{\beta _{{\text{st}}}}}}{{{\delta ^2}}}.
\end{equation}

It can be seen from (26) that ${\gamma _{\text{P}}}$ increases with the number of passive unit cells ${N_{\text{p}}}$. For the limited RIS power budget, the optimal number of passive unit cells can be expressed as
\begin{equation}
\label{eq-27}
N_{\text{p}}^{\text{*}} = \left\lfloor {\frac{{{P_{{\text{PRIS}}}}}}{{{P_{\text{e}}}}}} \right\rfloor ,
\end{equation}
where $\left\lfloor \cdot \right\rfloor$ means a rounding-down operation. At this time, the optimal SNR can be expressed as
\begin{equation}
\label{eq-28}
\gamma _{\text{P}}^{\text{*}} = \frac{{pP_{{\text{PRIS}}}^2N_{\text{t}}^2{\beta _{{\text{rd}}}}{\beta _{{\text{tr}}}}{\beta _{{\text{st}}}}}}{{P_{\text{e}}^2{\delta ^2}}}.
\end{equation}

\subsection{ARIS-assisted transmission scenario}
Similar to PRIS, in ARIS-assisted transmission scenario, our goal is to maximize the UE SNR by optimizing the phase shift matrices of the transmission RIS and ARIS $\left\{ {{\bm{\Phi} _{\text{t}}},{\bm{\Phi} _{\text{a}}}} \right\}$ and the amplification factor $\beta $ of the ARIS. The formula for maximizing the UE SNR is expressed as follows
\begin{align}
(\text{\bf{P3}}): \mathop {\max }\limits_{{\bm{\Phi} _{\text{t}}},{\bm{\Phi} _{\text{p}}},\beta } &\frac{{p{{\left| {\bm{h}{{_{{\text{rd}}}^{\text{a}}}^T}\beta {\bm{\Phi} _{\text{a}}}\bm{h}_{{\text{tr}}}^{\text{a}}{\bm{\Phi} _{\text{t}}}{\bm{h}_{{\text{st}}}}} \right|}^2}}}{{\left( {{{\left\| {\bm{h}{{_{{\text{rd}}}^{\text{a}}}^T}\beta {\bm{\Phi} _{\text{a}}}} \right\|}^2} + 1} \right){\delta ^2}}},\label{eq-29}\\
\text{s.t.} &\left| {{{\left[ {{\bm{\Phi} _{\text{t}}}} \right]}_i}} \right| = 1,\forall i \in {{\mathcal N}_{\text{t}}},\tag{\ref{eq-29}{a}} \label{eq-29a}\\
&\left| {{{\left[ {{\bm{\Phi} _{\text{a}}}} \right]}_j}} \right| = 1,\forall j \in {{\mathcal N}_{\text{a}}},\tag{\ref{eq-29}{b}} \label{eq-29b}\\
&\beta  \leqslant {\beta _{\max }},\tag{\ref{eq-29}{c}} \label{eq-29c}\\
&{\beta ^2}\left( {p{{\left\| {{\bm{\Phi} _{\text{a}}}\bm{h}_{{\text{tr}}}^{\text{a}}{\bm{\Phi} _{\text{t}}}{\bm{h}_{{\text{st}}}}} \right\|}^2}{\text{ + }}{\delta ^2}{{\left\| {{\bm{\Phi} _{\text{a}}}{\bm{I}_{{N_{\text{a}}}}}} \right\|}^2}} \right) \notag \\
&\leqslant \left( {{P_{{\text{ARIS}}}} - {N_{\text{a}}}\left( {{P_{\text{e}}} + {P_{{\text{DC}}}}} \right)} \right)v.\tag{\ref{eq-29}{d}} \label{eq-29d}
\end{align}

For a fair comparison, this paper assumes that the ARIS location and the PRIS location are the same. According to the geometric relationship between transmission RIS and ARIS, the channel matrix $\bm{h}_{{\text{tr}}}^{\text{a}}$ can also be decomposed into the product of two eigenvectors ${\bm{g}_{\text{t}}}$ and ${\bm{g}_{\text{a}}}$, namely
\begin{equation}
\label{eq-30}
\bm{h}_{{\text{tr}}}^{\text{a}} = q{\bm{g}_{\text{a}}}\bm{g}_{\text{t}}^T,
\end{equation}
where ${\bm{g}_{\text{a}}} \in {{\mathbb C}^{{N_{\text{a}}}}}$ denotes element-wise reflection vectors at ARIS. Using (30), the CSI from transmission RIS to ARIS can be calculated according to the geometric relationship between them. Similar to PRIS, the reflection coefficient matrix of ARIS is configured as follows
\begin{equation}
\label{eq-31}
\phi _j^{\text{a}} = {\left( {\frac{{{{\left[ {\bm{h}_{{\text{rd}}}^{\text{a}}} \right]}_j}{{\left[ {{\bm{g}_{\text{a}}}} \right]}_j}}}{{\left| {{{\left[ {\bm{h}_{{\text{rd}}}^{\text{a}}} \right]}_j}} \right|\left| {{{\left[ {{\bm{g}_{\text{a}}}} \right]}_j}} \right|}}} \right)^*},{\text{ }}\;j \in {{\mathcal N}_{\text{a}}}.
\end{equation}

According to the above results, the transmission coefficient matrix of transmission RIS and the reflection coefficient matrix of ARIS are substituted into (29), and (P3) can be rewritten as
\begin{align}
(\text{\bf{P4}}): \mathop {\max }\limits_\beta  &\frac{{p{\beta ^2}{{\left| {\sum\limits_{m = 1}^{{N_{\text{t}}}} {\sum\limits_{n = 1}^{{N_{\text{a}}}} {\left| {\bm{h}_{{\text{rd}}n}^{\text{a}}} \right|} \left| {\bm{h}_{{\text{tr}}nm}^{\text{a}}} \right|\left| {{\bm{h}_{{\text{st}}m}}} \right|} } \right|}^2}}}{{\left( {{\beta ^2}\sum\limits_{n = 1}^{{N_{\text{a}}}} {\left| {\bm{h}_{{\text{rd}}n}^{\text{a}}} \right|}  + 1} \right){\delta ^2}}},\label{eq-32}\\
\text{s.t.} & (\text{29c})-(\text{29d}).\notag
\end{align}

After a simplification like (26) and some simple mathematical derivation, (32) can be rewritten as
\begin{align}
(\text{\bf{P5}}): \mathop {\max }\limits_\beta  &\frac{{p{\beta ^2}N_{\text{t}}^2N_{\text{a}}^2{\beta _{{\text{rd}}}}{\beta _{{\text{tr}}}}{\beta _{{\text{st}}}}}}{{\left( {{\beta ^2}{\beta _{{\text{rd}}}}{N_{\text{a}}} + 1} \right){\delta ^2}}},\label{eq-33}\\
\text{s.t.} &  (\text{29c}),\notag \\
&\beta  \leqslant \sqrt {\frac{{\left( {{P_{{\text{ARIS}}}} - {N_{\text{a}}}\left( {{P_{\text{e}}} + {P_{{\text{DC}}}}} \right)} \right)v}}{{{N_{\text{a}}}\left( {p{\beta _{{\text{tr}}}}{\beta _{{\text{st}}}} + {\delta ^2}} \right)}}}.\tag{\ref{eq-33}{a}} \label{eq-33a}
\end{align}

(33) indicates that the objective function increases with $\beta $. When $\beta $ takes the following value, the optimal SNR can be obtained
\begin{equation}
\label{eq-34}
\beta  = \min \left\{ {\sqrt {\frac{{\left( {{P_{{\text{ARIS}}}} - {N_{\text{a}}}\left( {{P_{\text{e}}} + {P_{{\text{DC}}}}} \right)} \right)v}}{{{N_{\text{a}}}\left( {p{\beta _{{\text{tr}}}}{\beta _{{\text{st}}}} + {\delta ^2}} \right)}}} ,{\beta _{\max }}} \right\}.
\end{equation}

{\bf{Lemma 1}} From the above studies, it can be concluded that there is a trade-off between the number of ARIS active unit cells ${N_{\text{a}}}$ and the amplification factor $\beta $ under the power budget ${P_{{\text{ARIS}}}}$. To find the optimal SNR under the ARIS-assisted transmission scenario, the optimal number of ARIS unit cells can be expressed as a $\left\lfloor {N_{\text{a}}^{\text{*}}} \right\rfloor $ and $\left\lceil {N_{\text{a}}^{\text{*}}} \right\rceil $, where $\left\lceil \cdot \right\rceil$ denotes a rounding-up operation. $N_{\text{a}}^{\text{*}}$can be expressed as
\begin{equation}
\label{eq-35}
N_{\text{a}}^{\text{*}} = \max \left\{ {{N_{{\text{a}},1}},{N_{{\text{a}},2}}} \right\},
\end{equation}
\begin{equation}
\label{eq-36}
{N_{{\text{a}},1}} = \frac{{z - \sqrt {{z^2} - v{\beta _{{\text{rd}}}}{P_{{\text{ARIS}}}}z} }}{{v{\beta _{{\text{rd}}}}\left( {{P_{\text{e}}} + {P_{{\text{DC}}}}} \right)}},
\end{equation}
\begin{equation}
\label{eq-37}
{N_{{\text{a}},2}} = \frac{{{P_{{\text{ARIS}}}}v}}{{\beta _{\max }^2\left( {p{\beta _{{\text{tr}}}}{\beta _{{\text{st}}}} + {\delta ^2}} \right) + \left( {{P_{\text{e}}} + {P_{{\text{DC}}}}} \right)v}},
\end{equation}
where $z = p{\beta _{{\text{tr}}}}{\beta _{{\text{st}}}} + v{\beta _{{\text{rd}}}}{P_{{\text{ARIS}}}} + {\delta ^2}$.

{\bf{proof}}: First, when (33a) holds, there are ${N_{\text{a}}} \geqslant {N_{{\text{a}},2}}$, $\beta  = \sqrt {\frac{{\left( {{P_{{\text{ARIS}}}} - {N_{\text{a}}}\left( {{P_{\text{e}}} + {P_{{\text{DC}}}}} \right)} \right)v}}{{{N_{\text{a}}}\left( {p{\beta _{{\text{tr}}}}{\beta _{{\text{st}}}} + {\delta ^2}} \right)}}} $, and the optimal SNR is expressed as:
\begin{equation}
\label{eq-38}
{\gamma _{{\text{A,1}}}} = \frac{{pN_{\text{t}}^2{N_{\text{a}}}{\beta _{{\text{rd}}}}{\beta _{{\text{tr}}}}{\beta _{{\text{st}}}}v\left( {{P_{{\text{ARIS}}}} - {N_{\text{a}}}\left( {{P_{\text{e}}} + {P_{{\text{DC}}}}} \right)} \right)}}{{\left( {\left( {{P_{{\text{ARIS}}}} - {N_{\text{a}}}\left( {{P_{\text{e}}} + {P_{{\text{DC}}}}} \right)} \right){\beta _{{\text{rd}}}}v + p{\beta _{{\text{tr}}}}{\beta _{{\text{st}}}} + {\delta ^2}} \right){\delta ^2}}}.
\end{equation}

By taking the derivation of (38) to ${N_{\text{a}}}$, it can be seen that when ${N_{\text{a}}} < {N_{{\text{a}},1}}$, ${\gamma _{\text{A}}}$ increases; when ${N_{\text{a}}} \geqslant {N_{{\text{a}},1}}$, ${\gamma _{\text{A}}}$ decreases, and the optimal point is ${N_{\text{a}}} = {N_{{\text{a}},1}}$.

When (29c) holds, there are ${N_{\text{a}}} < {N_{{\text{a}},2}}$, $\beta  = {\beta _{\max }}$, and the optimal SNR is
\begin{equation}
\label{eq-39}
{\gamma _{{\text{A,2}}}} = \frac{{p\beta _{\max }^2N_{\text{t}}^2N_{\text{a}}^2{\beta _{{\text{rd}}}}{\beta _{{\text{tr}}}}{\beta _{{\text{st}}}}}}{{\left( {{\beta _{{\text{rd}}}}\beta _{\max }^2{N_{\text{a}}} + 1} \right){\delta ^2}}}.
\end{equation}

At this time, ${\gamma _{\text{A}}}$ increases with ${N_{\text{a}}}$, and the optimum point is ${N_{\text{a}}} = {N_{{\text{a}},2}}$. Here, this paper treats discrete variables ${N_{\text{a}}}$ as continuous variables. So, relative to ${N_{\text{a}}}$, the SNR function is also continuous, indicating that the value of the function does not have any abrupt changes.

When ${N_{{\text{a,1}}}} \leqslant {N_{{\text{a}},2}}$, ${N_{\text{a}}} < {N_{{\text{a}},2}}$, (29c) holds, and ${\gamma _{\text{A}}}$ increases with ${N_{\text{a}}}$; when ${N_{\text{a}}} \geqslant {N_{{\text{a}},2}}$, (33a) holds, and ${\gamma _{\text{A}}}$ decreases as ${N_{\text{a}}}$ increases. So, the optimal value is $N_{\text{a}}^{\text{*}} = {N_{{\text{a}},2}}$.

When ${N_{{\text{a,1}}}} > {N_{{\text{a}},2}}$, ${N_{\text{a}}} < {N_{{\text{a}},2}}$, (29c) holds, and ${\gamma _{\text{A}}}$ increases with ${N_{\text{a}}}$; when ${N_{{\text{a}},2}} \leqslant {N_{\text{a}}} < {N_{{\text{a}},1}}$, (33a) holds, and ${\gamma _{\text{A}}}$ increases with ${N_{\text{a}}}$; when ${N_{\text{a}}} \geqslant {N_{{\text{a}},1}}$, ${\gamma _{\text{A}}}$ decreases as ${N_{\text{a}}}$ increases. So, the optimal value is $N_{\text{a}}^{\text{*}} = {N_{{\text{a}},1}}$.

In summary, we have $N_{\text{a}}^{\text{*}} = \max \left\{ {{N_{{\text{a}},1}}, {N_{{\text{a}},2}}} \right\}$. Since ${N_{\text{a}}}$ is an integer, the optimal number of ARIS unit cells can be expressed as $\left\lfloor {N_{\text{a}}^{\text{*}}} \right\rfloor $ and $\left\lceil {N_{\text{a}}^{\text{*}}} \right\rceil $.

\subsection{HRIS-assisted transmission scenario}
In HRIS-assisted transmission scenario, our goal is also to maximize the UE SNR of the HRIS-assisted transmission scenario by optimizing the phase shift matrices $\left\{ {{\bm{\Phi} _{\text{t}}},{\bm{\Phi} _{{\text{ha}}}},{\bm{\Phi} _{{\text{hp}}}}} \right\}$ of the transmission RIS and HRIS and the amplification factor $\beta $ of the active unit cells of HRIS. The formula for maximizing the UE SNR is expressed as follows
\begin{align}
(\text{\bf{P6}}): \notag \\
    \mathop {\max }\limits_{{\bm{\Phi} _{\text{t}}},{\bm{\Phi} _{{\text{hp}}}},{\bm{\Phi _{{\text{ha}}}}},\beta} &\frac{{p{{\left| {\bm{h}{{_{{\text{rd}}}^{{\text{hp}}}}^T}{\bm{\Phi} _{{\text{hp}}}}\bm{h}_{{\text{tr}}}^{{\text{hp}}}{\bm{\Phi} _{\text{t}}}{\bm{h}_{{\text{st}}}} + \bm{h}{{_{{\text{rd}}}^{{\text{ha}}}}^T}\beta {\bm{\Phi} _{{\text{ha}}}}\bm{h}_{{\text{tr}}}^{{\text{ha}}}{\bm{\Phi} _{\text{t}}}{\bm{h}_{{\text{st}}}}} \right|}^2}}}{{\left( {{{\left\| {\bm{h}{{_{{\text{rd}}}^{{\text{ha}}}}^T}\beta {\bm{\Phi} _{{\text{ha}}}}} \right\|}^2} + 1} \right){\delta ^2}}},\label{eq-40}\\
\text{s.t.} &\left| {{{\left[ {{\bm{\Phi} _{\text{t}}}} \right]}_i}} \right| = 1,\forall i \in {{\mathcal N}_{\text{t}}},\tag{\ref{eq-40}{a}} \label{eq-40a}\\
&\left| {{{\left[ {{\bm{\Phi} _{{\text{ha}}}}} \right]}_j}} \right| = 1,\forall j \in {{\mathcal N}_{{\text{ha}}}},\tag{\ref{eq-40}{b}} \label{eq-40b}\\
&\left| {{{\left[ {{\bm{\Phi} _{{\text{hp}}}}} \right]}_z}} \right| = 1,\forall z \in {{\mathcal N}_{{\text{hp}}}},\tag{\ref{eq-40}{c}} \label{eq-40c}\\
&\beta  \leqslant {\beta _{\max }},\tag{\ref{eq-40}{d}} \label{eq-40d}\\
&{\beta ^2}\left( {p{{\left\| {{\bm{\Phi} _{{\text{ha}}}}\bm{h}_{{\text{tr}}}^{{\text{ha}}}{\bm{\Phi} _{\text{t}}}{\bm{h}_{{\text{st}}}}} \right\|}^2}{\text{ + }}{\delta ^2}{{\left\| {{\bm{\Phi} _{{\text{ha}}}}{\bm{I}_{{N_{{\text{ha}}}}}}} \right\|}^2}} \right) \notag \\ 
&\leqslant \left( {{P_{{\text{HRIS}}}} - {N_{\text{h}}}{P_{\text{e}}} - {N_{{\text{ha}}}}{P_{{\text{DC}}}}} \right)v .\tag{\ref{eq-40}{e}}
\label{eq-40e}
\end{align}

Similarly, this paper assumes that the HRIS location is the same as the PRIS and ARIS. According to the geometric relationship of transmission RIS and HRIS, the channel matrix $\bm{h}_{{\text{tr}}}^{{\text{ha}}}$ can be decomposed into the product of two eigenvectors ${\bm{g}_{\text{t}}}$ and ${\bm{g}_{{\text{ha}}}}$, namely
\begin{equation}
\label{eq-41}
\bm{h}_{{\text{tr}}}^{{\text{ha}}} = q{\bm{g}_{{\text{ha}}}}\bm{g}_{\text{t}}^T,
\end{equation}
where ${\bm{g}_{{\text{ha}}}} \in {{\mathbb C}^{{N_{{\text{ha}}}}}}$ denotes an element-wise reflection vector at active unit cells in HRIS. The channel matrix $\bm{h}_{{\text{tr}}}^{{\text{hp}}}$ is decomposed into the product of two eigenvectors ${\bm{g}_{\text{t}}}$ and $\bm{g}_{\text{hp}}$, namely
\begin{equation}
\label{eq-42}
\bm{h}_{{\text{tr}}}^{{\text{hp}}} = q{\bm{g}_{{\text{hp}}}}\bm{g}_{\text{t}}^T,
\end{equation}
where ${\bm{g}_{{\text{hp}}}} \in {{\mathbb C}^{{N_{{\text{hp}}}}}}$ denotes an element-wise reflection vector at passive unit cells in HRIS. Using (41) and (42), the CSI from transmission RIS to HRIS can be calculated based on the geometric relationship between them, and the reflection coefficient matrix of HRIS can be configured as follows
\begin{equation}
\label{eq-43}
\phi _j^{{\text{ha}}} = {\left( {\frac{{{{\left[ {\bm{h}_{{\text{rd}}}^{{\text{ha}}}} \right]}_j}{{\left[ {{\bm{g}_{{\text{ha}}}}} \right]}_j}}}{{\left| {{{\left[ {\bm{h}_{{\text{rd}}}^{{\text{ha}}}} \right]}_j}} \right|\left| {{{\left[ {{\bm{g}_{{\text{ha}}}}} \right]}_j}} \right|}}} \right)^*},{\text{ }}\;j \in {{\mathcal N}_{{\text{ha}}}},
\end{equation}
\begin{equation}
\label{eq-44}
\phi _z^{{\text{hp}}} = {\left( {\frac{{{{\left[ {\bm{h}_{{\text{rd}}}^{{\text{hp}}}} \right]}_z}{{\left[ {{\bm{g}_{{\text{hp}}}}} \right]}_z}}}{{\left| {{{\left[ {\bm{h}_{{\text{rd}}}^{{\text{hp}}}} \right]}_z}} \right|\left| {{{\left[ {{\bm{g}_{{\text{hp}}}}} \right]}_z}} \right|}}} \right)^*},{\text{ }}\;z \in {{\mathcal N}_{{\text{hp}}}}.
\end{equation}

According to the above results, the transmission RIS transmission coefficient matrix and the HRIS reflection coefficient matrix are substituted into (40), and (P6) is transformed into the following form
\begin{align}
(\text{\bf{P7}}): \mathop {\max }\limits_\beta  &\frac{{p{{\left| {\beta {C_{{\text{ha}}}}{\text{ + }}{C_{{\text{hp}}}}} \right|}^2}}}{{\left( {{\beta ^2}\sum\limits_{n = 1}^{{N_{{\text{ha}}}}} {\left| {\bm{h}_{{\text{rd}}n}^{{\text{ha}}}} \right|}  + 1} \right){\delta ^2}}},\label{eq-45}\\
\text{s.t.} & (\text{40d})-(\text{40e}),\notag 
\end{align}
where $C_{\text{ha}} = \sum\limits_{m = 1}^{{N_{\text{t}}}} {\sum\limits_{n = 1}^{{N_{{\text{ha}}}}} {\left| {\bm{h}_{{\text{rd}}n}^{{\text{ha}}}} \right|} \left| {\bm{h}_{{\text{tr}}nm}^{{\text{ha}}}} \right|\left| {{\bm{h}_{{\text{st}}m}}} \right|}$ and $C_{\text{hp}} = {\sum\limits_{m = 1}^{{N_{\text{t}}}} {\sum\limits_{n = 1}^{{N_{{\text{hp}}}}} {\left| {\bm{h}_{{\text{rd}}n}^{{\text{hp}}}} \right|} \left| {\bm{h}_{{\text{tr}}nm}^{{\text{hp}}}} \right|\left| {{\bm{h}_{{\text{st}}m}}} \right|} }$.

After a similar simplification of (26) and some simple mathematical simplifications, (45) can be rewritten as
\begin{align}
(\text{\bf{P8}}): \mathop {\max }\limits_\beta  &\frac{{p{N_{\text{t}}}^2{\beta _{{\text{rd}}}}{\beta _{{\text{tr}}}}{\beta _{{\text{st}}}}{{\left( {{N_{\text{h}}} + {N_{{\text{ha}}}}\left( {\beta  - 1} \right)} \right)}^2}}}{{\left( {{\beta ^2}{N_{{\text{ha}}}}{\beta _{{\text{rd}}}} + 1} \right){\delta ^2}}},\label{eq-46}\\
\text{s.t.} & (\text{40d}),\notag \\
&\beta  \leqslant \sqrt {\frac{{\left( {{P_{{\text{HRIS}}}} - {N_{\text{h}}}{P_{\text{e}}} - {N_{{\text{ha}}}}{P_{{\text{DC}}}}} \right)v}}{{{N_{{\text{ha}}}}\left( {p{\beta _{{\text{tr}}}}{\beta _{{\text{st}}}} + {\delta ^2}} \right)}}}.\tag{\ref{eq-46}{a}} \label{eq-46a}
\end{align}

(46) indicates that the objective function increases with $\beta $. When $\beta $ takes the following value, the optimal SNR can be obtained
\begin{equation}
\label{eq-47}
\beta  = \min \left\{ {\sqrt {\frac{{\left( {{P_{{\text{HRIS}}}} - {N_{\text{h}}}{P_{\text{e}}} - {N_{{\text{ha}}}}{P_{{\text{DC}}}}} \right)v}}{{{N_{{\text{ha}}}}\left( {p{\beta _{{\text{tr}}}}{\beta _{{\text{st}}}} + {\delta ^2}} \right)}}} ,{\beta _{\max }}} \right\}.
\end{equation}

Since the value of (46) increases as ${N_{{\text{ha}}}}$ increases, there is no optimal ${N_{{\text{ha}}}}$. So, this paper finds the optimal SNR given the number of active unit cells in HRIS. When (46a) holds, $\beta  = \sqrt {\frac{{\left( {{P_{{\text{HRIS}}}} - {N_{\text{h}}}{P_{\text{e}}} - {N_{{\text{ha}}}}{P_{{\text{DC}}}}} \right)v}}{{{N_{{\text{ha}}}}\left( {p{\beta _{{\text{tr}}}}{\beta _{{\text{st}}}} + {\delta ^2}} \right)}}} $, the optimal SNR can be expressed as:
\begin{align}
\label{eq-48}
&{\gamma _{{\text{H,1}}}} = \notag  \\
&\frac{{p{N_{\text{t}}}^2{\beta _{{\text{rd}}}}{\beta _{{\text{tr}}}}{\beta _{{\text{st}}}}{{\left( {\left( {{N_{\text{h}}} - {N_{{\text{ha}}}}} \right)\sqrt {p{\beta _{{\text{tr}}}}{\beta _{{\text{st}}}} + {\delta ^2}}  + \sqrt {{N_{{\text{ha}}}}v{P_{\text{h}}}} } \right)}^2}}}{{\left( {v{\beta _{{\text{rd}}}}{P_{\text{h}}} + p{\beta _{{\text{tr}}}}{\beta _{{\text{st}}}} + {\delta ^2}} \right){\delta ^2}}}.
\end{align}

When (40d) holds, there is $\beta  = {\beta _{\max }}$, and the optimal SNR at this time is
\begin{equation}
\label{eq-49}
{\gamma _{{\text{H,2}}}} = \frac{{p{N_{\text{t}}}^2{\beta _{{\text{rd}}}}{\beta _{{\text{tr}}}}{\beta _{{\text{st}}}}{{\left( {{N_{\text{h}}} + {N_{{\text{ha}}}}\left( {{\beta _{\max }} - 1} \right)} \right)}^2}}}{{\left( {{\beta _{{\text{rd}}}}\beta _{\max }^2{N_{{\text{ha}}}} + 1} \right){\delta ^2}}}.
\end{equation}

\section{Performance Analysis}
For the downlink system, this paper compares the achievable data rate, energy efficiency, and the relationship between energy efficiency and spectral efficiency of PRIS, ARIS, and HRIS-assisted transmission scenarios in mmWave indoor signal enhancement.

\subsection{Achievable rate}
According to the above analysis, the achievable rates of PRIS, ARIS, and HRIS-assisted transmission scenarios to the mmWave indoor signal enhancement scheme can be expressed as
  \begin{flalign}
& {R^{\text{P}}} = {\log _2}\left( {1 + \frac{{pP_{{\text{PRIS}}}^2N_{\text{t}}^2{\beta _{{\text{rd}}}}{\beta _{{\text{tr}}}}{\beta _{{\text{st}}}}}}{{P_{\text{e}}^2{\delta ^2}}}} \right),&
\end{flalign}
\label{eq-50}
 \begin{flalign}
\label{eq-51}
&{R^{\text{A}}} = \left\{ 
\begin{aligned}
   {{\log }_2}\left( {1 + {\gamma _{{\text{A,1}}}}} \right),&\beta  = \sqrt {\frac{{\left( {{P_{{\text{ARIS}}}} - {N_{\text{a}}}\left( {{P_{\text{e}}} + {P_{{\text{DC}}}}} \right)} \right)v}}{{{N_{\text{a}}}\left( {p{\beta _{{\text{tr}}}}{\beta _{{\text{st}}}} + {\delta ^2}} \right)}}}   \\ 
   {{\log }_2}\left( {1 + {\gamma _{{\text{A,2}}}}} \right),&\beta  = {\beta _{\max }}   
\end{aligned}
\right. , &
 \end{flalign}
\begin{equation}
\label{eq-52}
{R^{\text{H}}} = \left\{ 
\begin{aligned}
  {{\log }_2}\left( {1 + {\gamma _{{\text{H,1}}}}} \right),&\beta  = \sqrt {\frac{{\left( {{P_{{\text{HRIS}}}} - {N_{\text{h}}}{P_{\text{e}}} - {N_{{\text{ha}}}}{P_{{\text{DC}}}}} \right)v}}{{{N_{{\text{ha}}}}\left( {p{\beta _{{\text{tr}}}}{\beta _{{\text{st}}}} + {\delta ^2}} \right)}}} \\ 
  {{\log }_2}\left( {1 + {\gamma _{{\text{H,2}}}}} \right),&\beta  = {\beta _{\max }}  
\end{aligned}
\right. .
\end{equation}
\subsection{Energy efficiency}
First, consider the total power of the system. The total power ${P_{{\text{total}}}}$ of the system includes the transmit power and the dissipated power of the hardware. In PRIS-assisted transmission scenario, the total power can be expressed as
\begin{equation}
\label{eq-53}
P_{{\text{total}}}^{\text{P}} = \frac{p}{v} + {P_{\text{s}}} + {P_{\text{d}}} + {N_{\text{t}}}{P_{\text{e}}} + {P_{{\text{PRIS}}}},
\end{equation}
where ${P_{\text{s}}}$ and ${P_{\text{d}}}$ represent the hardware dissipated power of the BS and UE, respectively. In ARIS-assisted transmission scenario, the total power can be expressed as
\begin{equation}
\label{eq-54}
P_{{\text{total}}}^{\text{A}} = \frac{p}{v} + {P_{\text{s}}} + {P_{\text{d}}} + {N_{\text{t}}}{P_{\text{e}}} + {P_{{\text{ARIS}}}},
\end{equation}
In HRIS-assisted transmission scenario, the total power can be expressed as
\begin{equation}
\label{eq-55}
P_{{\text{total}}}^{\text{H}} = \frac{p}{v} + {P_{\text{s}}} + {P_{\text{d}}} + {N_{\text{t}}}{P_{\text{e}}} + {P_{{\text{HRIS}}}}.
\end{equation}

With the above definition of total power, the expression for energy efficiency can be obtained from [38]
\begin{equation}
\label{eq-56}
EE = \frac{R}{{{P_{{\text{total}}}}}}.
\end{equation}
\subsection{Trade-off between spectral efficiency and energy efficiency}
Since the spectral efficiency, i.e., the data rate is independent of the total power, the energy efficiency is related to the total power. To investigate the relationship between spectral efficiency and energy efficiency of the system, this paper adopts the following performance index (PI) to link the relationship between spectral efficiency and energy efficiency. PI can be expressed as
\begin{equation}
\label{eq-57}
{\text{PI}} = {R^\lambda } + {\left( {\frac{R}{{{P_{{\text{total}}}}}}} \right)^{1 - \lambda }}.
\end{equation}
where $\lambda $ is the adjustment coefficient of spectral efficiency and energy efficiency.

\section{Simulation Results}
\begin{figure}[!t]
\centering
\includegraphics[width=3.2in]{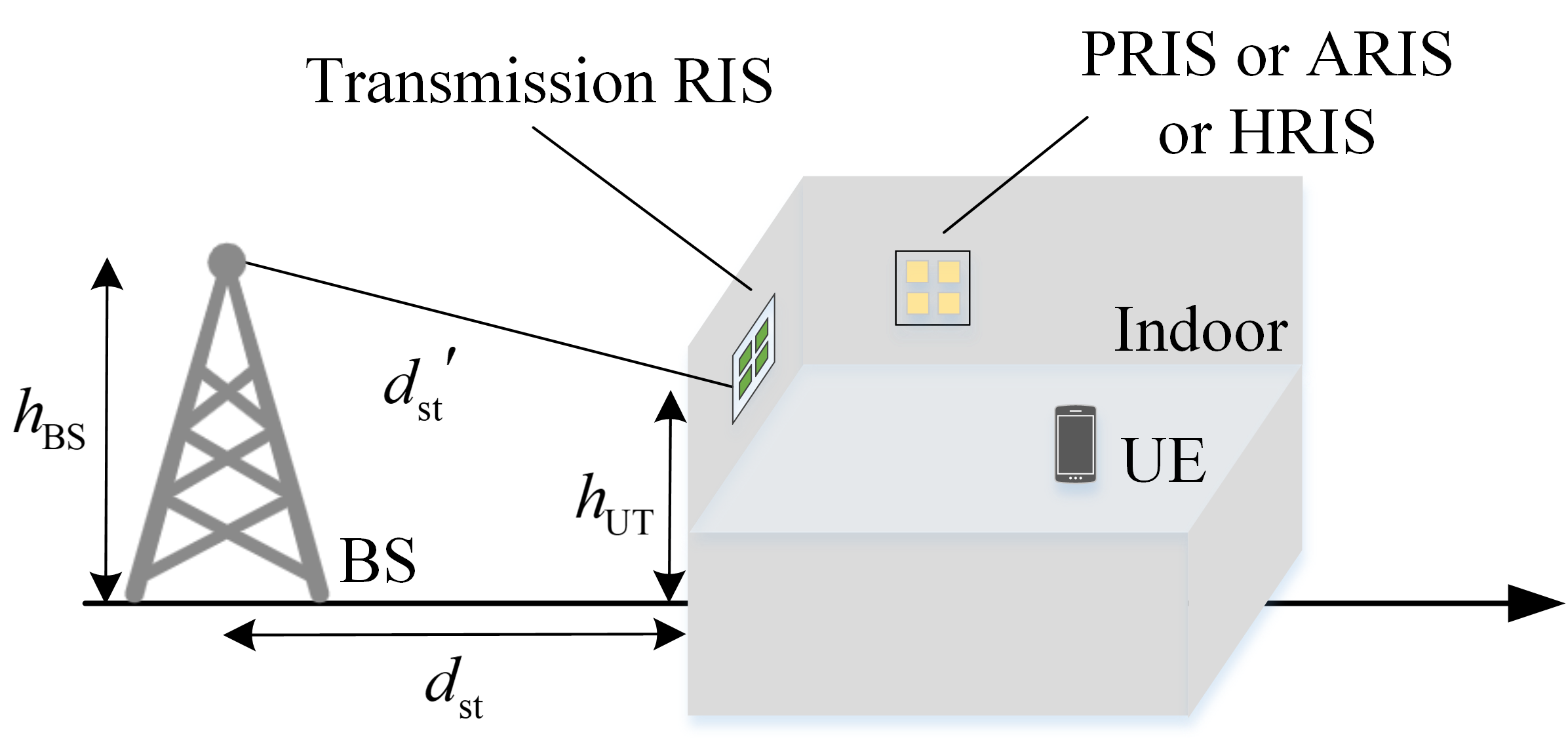}
\caption{Simulation diagram.}
\label{fig02}
\end{figure}

This section compares the achievable rate, energy efficiency, and the relationship between energy efficiency and spectral efficiency of PRIS, ARIS, and HRIS-assisted transmission scenarios in the mmWave indoor signal enhancement scheme. The scenario shown in Fig.~\ref{fig02} is considered, and a three-dimensional Cartesian coordinate is used, where the height of BS is ${h_{{\text{BS}}}}$, the heights of transmission RIS, PRIS, ARIS, HRIS, and UE are ${h_{{\text{UT}}}}$, the horizontal distance from the BS to transmission RIS is ${d_{{\text{st}}}}$, and the distance from the BS to transmission RIS is ${d_{{\text{st}}}}^\prime $. Other simulation parameters are presented in Tab. I.
\begin{table}[!t]
\caption{Simulation Parameters\label{tab1}}
\setlength\tabcolsep{16pt}
\renewcommand{\arraystretch}{1.2} 
\centering
\begin{tabular}{|c|c|}
\hline
{\bf{Parameters}} & {\bf{Values}}\\
\hline
${h_{{\text{BS}}}}$ & 25 m\\
\hline
${h_{{\text{UT}}}}$ & 19.5 m\\
\hline
BS & (${d_{{\text{st}}}}$ m, 0 m, 19.5 m)\\
\hline
Transmission RIS & (${d_{{\text{st}}}}$ m, 0 m, 19.5 m)\\
\hline
PRIS, ARIS and HRIS & (${d_{{\text{st}}}}$ + 2 m, 4 m, 19.5 m)\\
\hline
UE & (${d_{{\text{st}}}}$ + 4 m , 2 m , 19.5 m)\\
\hline
\end{tabular}
\end{table}

The indoor channel model adopts the 3GPP InH Office path loss model~\cite{ref40} with a carrier frequency of ${f_c} =$ 28 GHz. Based on this, the mmWave indoor office path loss model can be represented as follows
\begin{flalign}
\label{eq-58}
  & {\beta _{{\text{InH}}}}({f_c},d)    & \notag \\
  & = {G_{\text{t}}} - 32.4 - 17.3{\log _{10}}\left( d \right) - 20{\log _{10}}\left( {{f_c}} \right) - {\sigma _{{\text{InH}}}},
\end{flalign}
where ${G_{\text{t}}}$ is the BS transmit antenna gain, and ${\sigma _{{\text{InH}}}}$ is the shadow fading. The BS to transmission RIS channel adopts the LOS link of the 3GPP Urban Macro (UMa) path loss model~\cite{ref40}. The expression is
\begin{equation}
\label{eq-59}
\beta _{{\text{LOS}}}^{{\text{UMa}}}({f_c},{d_{{\text{st}}}}^\prime ) = {G_{\text{t}}} - \left\{ 
\begin{aligned}
 {\text{P}}{{\text{L}}_1} + \sigma _{{\text{LOS}}}^{{\text{UMa}}},& 10{\text{ m}} \leqslant {d_{{\text{st}}}} \leqslant {d_{{\text{bp}}}}  \\ 
 {\text{P}}{{\text{L}}_2} + \sigma _{{\text{LOS}}}^{{\text{UMa}}},& {d_{{\text{bp}}}} \leqslant {d_{{\text{st}}}} \leqslant 5{\text{ km}} 
\end{aligned}
\right. ,
\end{equation}
\begin{equation}
\label{eq-60}
{\text{P}}{{\text{L}}_1} = 28.0 + 22{\log _{10}}\left( {{d_{{\text{st}}}}^\prime } \right) + 20{\log _{10}}\left( {{f_c}} \right),
\end{equation}
\begin{align}
\label{eq-61}
{\text{P}}{{\text{L}}_2} = &28.0 + 40{\log _{10}}\left( {{d_{{\text{st}}}}^\prime } \right) + 20{\log _{10}}\left( {{f_c}} \right)  \notag \\ 
  &- 9{\log _{10}}\left( {{{\left( {{d_{{\text{bp}}}}} \right)}^2} + {{\left( {{h_{{\text{BS}}}} - {h_{{\text{UT}}}}} \right)}^2}} \right) ,
\end{align}
where ${d_{{\text{bp}}}}$ is the breakpoint distance, and $\sigma _{{\text{LOS}}}^{{\text{UMa}}}$ is the shadow fading.

Suppose that the BS has an antenna gain of 5 dBi, and the UE antenna is an omnidirectional antenna with an antenna gain of 0 dBi. Using the 3GPP InH Office LOS channel fading model, transmission RIS to PRIS, ARIS, HRIS and PRIS, ARIS, HRIS to UE have LOS channels. Since a deterministic channel is used, shadow fading is not considered here.

\begin{figure}[!t]
\centering
\includegraphics[width=3.2in]{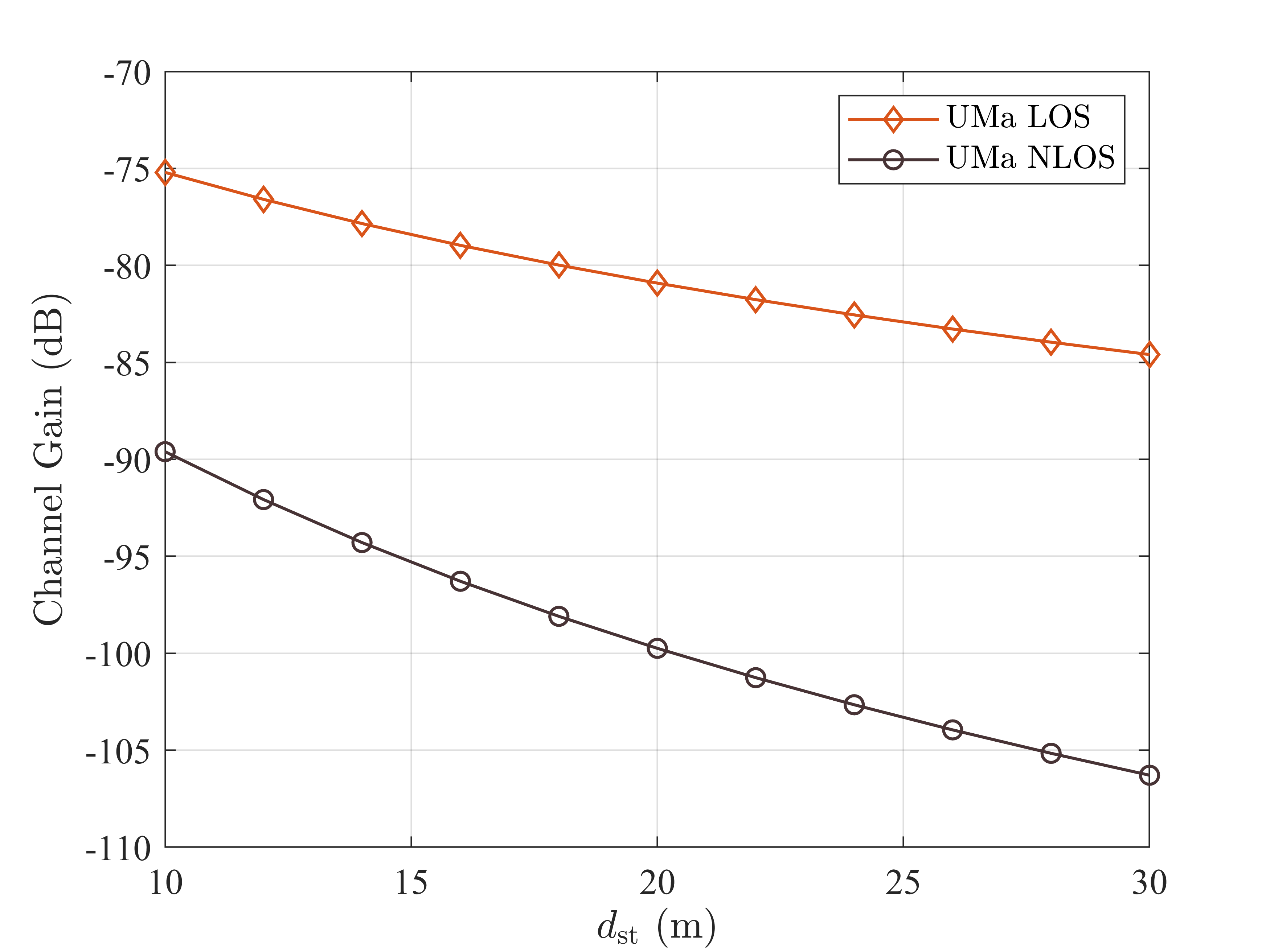}
\caption{The variation of the 3GPP UMa channel gain with distance ${d_{{\text{st}}}}$, where ${G_{\text{t}}} = 5{\text{ dBi}}$.}
\label{fig03}
\end{figure}

Fig.~\ref{fig03} shows the variation of the 3GPP UMa channel gain with distance ${d_{{\text{st}}}}$. It can be seen from the figure that in mmWave, the channel attenuation is extreme, and when ${d_{{\text{st}}}} =$ 10 m, the UMa LOS channel gain reaches -75 dB, and it increases gradually with ${d_{{\text{st}}}}$. The channel gain difference between the UMa non line-of-sight (NLOS) path and the UMa LOS path is -25 dB when ${d_{{\text{st}}}} =$ 10 m, and it increases gradually with ${d_{{\text{st}}}}$.

\begin{figure}[!t]
\centering
\includegraphics[width=3.2in]{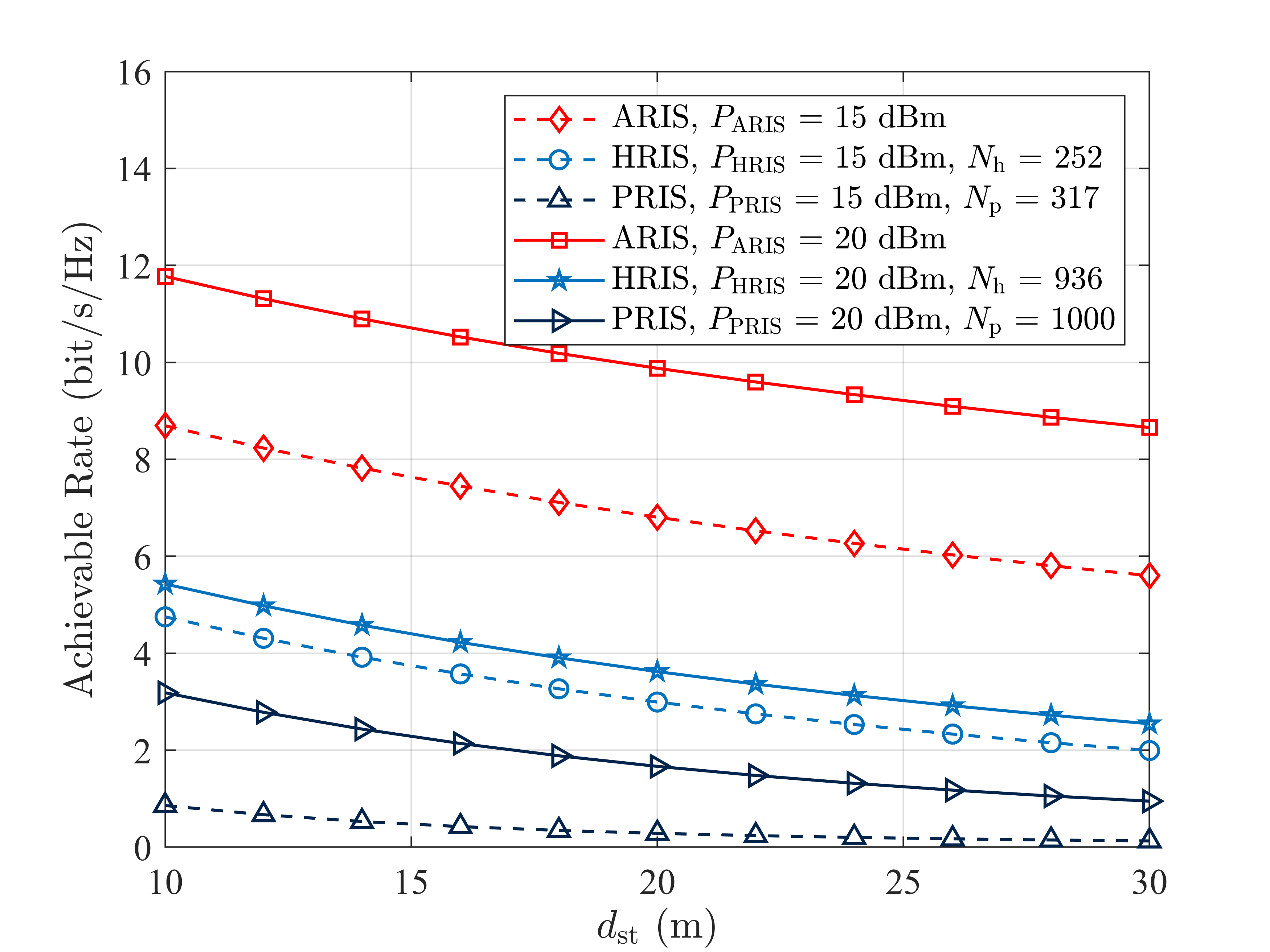}
\caption{Achievable rates under different ${d_{{\text{st}}}}$.}
\label{fig04}
\end{figure}

Fig.~\ref{fig04} shows the achievable data rates under different ${d_{{\text{st}}}}$. The simulation parameters are $p =$ 20 dBW with a bandwidth of $B =$ 10 MHz and a noise power of -94 dBm, ${N_{\text{t}}} =$ 400, ${N_{{\text{ha}}}} =$ 20, $\beta _{\max }^2 =$ 40 dB, ${P_{{\text{ARIS}}}} = {P_{{\text{HRIS}}}} = {P_{{\text{PRIS}}}} =$ 20 and 15 dBm, ${P_{\text{e}}} =$ -10 dBm, ${P_{{\text{DC}}}} =$ -5 dBm, and $v =$ 0.5.

As shown in Fig.~\ref{fig04}, as ${d_{{\text{st}}}}$ increases, the achievable rates under PRIS, ARIS, and HRIS-assisted transmission scenarios all decrease. Under the same amplification power budget, ARIS-assisted transmission scenario achieves the highest data rate and the smallest number of unit cells but uses more power amplifiers and increases the hardware cost. When the amplification power budget is increased, the achievable data rates of the three assisted transmission scenarios increase. This is because increasing the amplification power budget enables more power to be used to amplify the incident signal or use more unit cells, thereby increasing the achievable data rates. Under the same ${d_{{\text{st}}}}$ and amplification power budget, the achievable data rate under HRIS-assisted transmission scenario is more significant than that under PRIS-assisted transmission scenario, and the number of unit cells is smaller. For example, when ${P_{{\text{HRIS}}}} = {P_{{\text{PRIS}}}} =$ 15 dBm, ${d_{{\text{st}}}} =$ 12 m, the achievable data rate under HRIS-assisted transmission scenario is 84.5\% higher than that under PRIS-assisted transmission scenario (an increase of 3.64 bit/s/Hz), and the number of active unit cells ${N_{{\text{ha}}}}$ in HRIS is only 20; when ${P_{{\text{HRIS}}}} = {P_{{\text{PRIS}}}} =$ 20 dBm, ${d_{{\text{st}}}} =$ 12 m, the achievable data rate under HRIS-assisted transmission scenario is increased by 44.2\% (2.2 bit/s/Hz) compared with that under PRIS-assisted transmission scenario. This indicates that when only 20 active unit cells are used, the achievable data rate under HRIS-assisted transmission scenario is significantly improved compared to that under PRIS-assisted transmission scenario, which balances the hardware cost and data rate.

\begin{figure}[!t]
\centering
\includegraphics[width=3.2in]{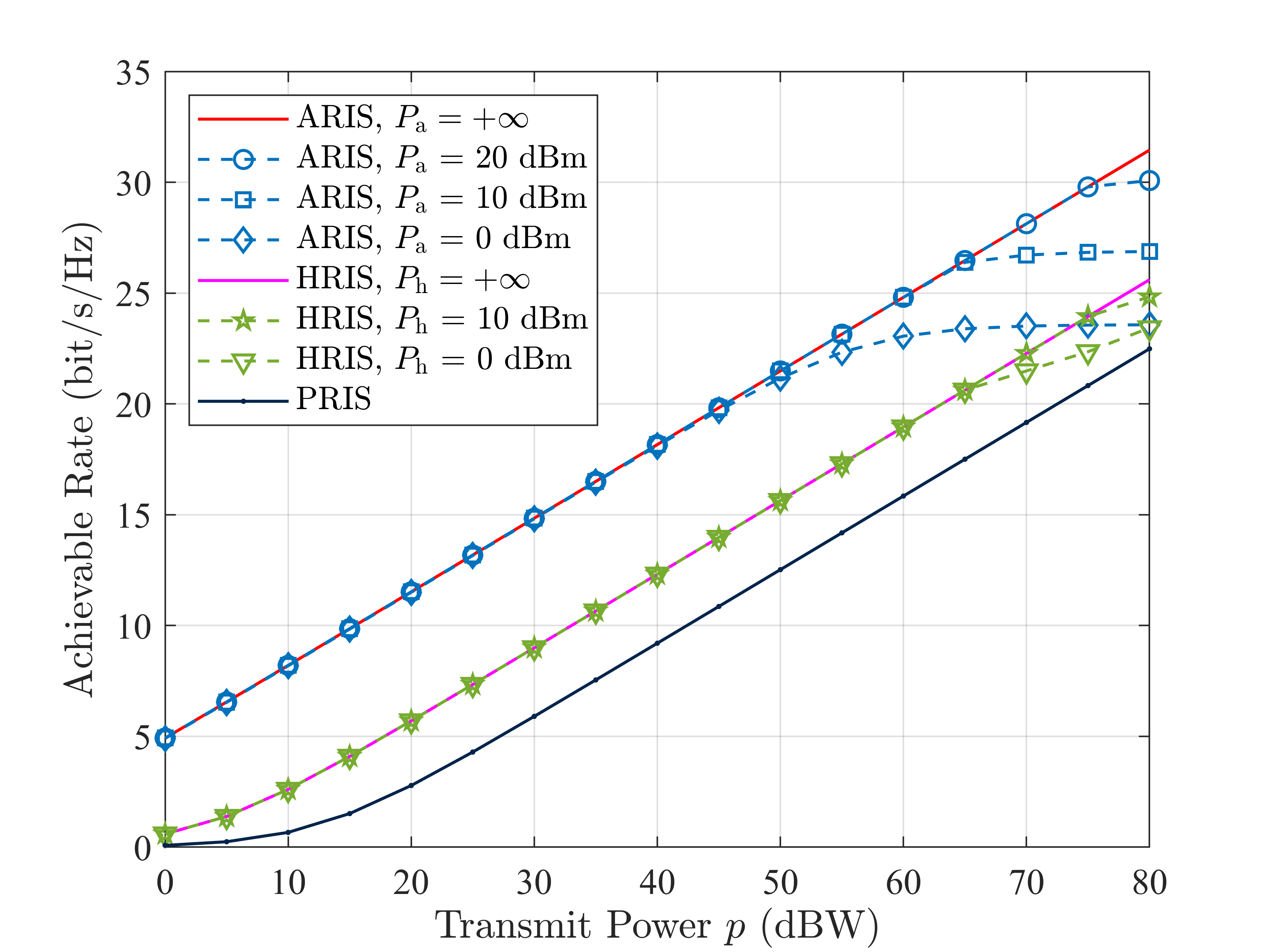}
\caption{Relationship between transmit power and achievable rate.}
\label{fig05}
\end{figure}

Then, this paper discusses the relationship between the transmit power $p$ and the achievable data rate, as shown in Fig.~\ref{fig05}. The simulation parameters are ${d_{{\text{st}}}} = $ 12 m, ${N_{\text{t}}}=$ 400, ${N_{\text{a}}} =$ 256, ${N_{\text{h}}} = {N_{\text{p}}} =$ 1000, ${N_{{\text{ha}}}} =$ 20, $\beta _{\max }^2 =$ 40 dB, ${P_{\text{a}}} =$ 20, 10 and 0 dBm, ${P_{\text{h}}} =$ 10 and 0 dBm with a bandwidth of $B =$ 10 MHz corresponding to a noise power of -94 dBm.

As shown in Fig.~\ref{fig05}, as $p$ increases, the achievable data rates under PRIS, ARIS, and HRIS-assisted transmission scenarios all increase, and the achievable data rate of ARIS is the largest. In general, under different parameter settings, ARIS-assisted transmission scenario performs better than PRIS-assisted transmission scenario because the unit cells in ARIS can amplify the incident signal, and the number of unit cells in ARIS is only about a quarter of that in PRIS, which takes up less volume and area. Under the condition of $\beta _{\max }^2 =$ 40 dB, it can be observed that when the transmit power is small, the ARIS-assisted transmission scenario has almost the same achievable data rate under different amplification power budgets. This indicates that amplification power budget constraints are ineffective in the case of a weak transmit power. However, as the transmit power increases, the blue dashed line does not change much, and the increase of the achievable data rate becomes moderate. According to the power budget constraint (33a), this constraint holds as the transmit power increases, leading to a decrease in the amplification gain of the ARIS-assisted transmission scenario. Therefore, the achievable data rate gain provided by ARIS is limited. Meanwhile, if the ARIS has a larger amplification power budget, the ARIS-assisted transmission scenario will perform better because it can provide more amplification gain under the same transmit power. Therefore, when designing the ARIS-assisted transmission scenario, its amplification gain must be reconfigured according to the transmit power.

The same phenomenon exists in the HRIS-assisted transmission scenario. Under a small transmit power, different amplification power budgets have almost the same achievable data rate. When the transmit power is considerable, the power budget constraint (46a) holds, which reduces the amplification gain and slows down the growth of the data rate. Meanwhile, the HRIS has a larger amplification power budget, and the HRIS-assisted transmission scenario performs better.
\begin{figure}[!t]
\centering
\includegraphics[width=3.2in]{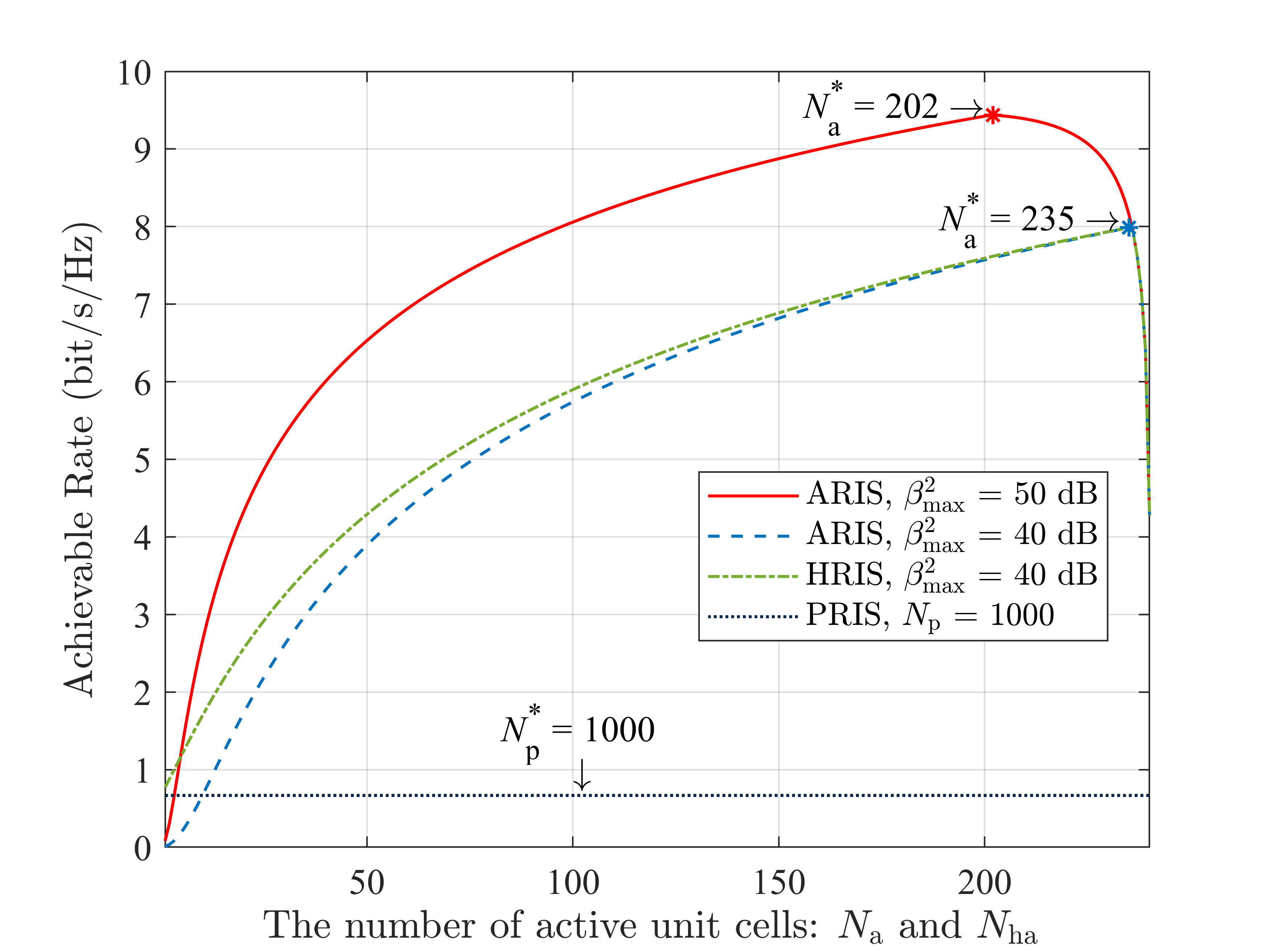}
\caption{The number of active unit cells versus the achievable rate.}
\label{fig06}
\end{figure}

Fig.~\ref{fig06} illustrates the relationship between the number of active unit cells and the achievable data rate. The simulation parameters are ${d_{{\text{st}}}} =$ 12 m, ${N_{\text{t}}}=$ 400, $\beta _{\max }^2 =$ 40 and 50 dB, ${P_{{\text{ARIS}}}} = {P_{{\text{HRIS}}}} = {P_{{\text{PRIS}}}} =$ 20 dBm, ${P_{\text{e}}} =$ -10 dBm, ${P_{{\text{DC}}}} =$ -5 dBm, $v =$ 0.5, with a bandwidth of $B =$ 10 MHz, corresponding to a noise power of -94 dBm.

Under a power budget of 20 dBm, PRIS can use 1000 unit cells. However, even with 1000 unit cells, the achievable data rate of the PRIS-assisted transmission scenario is still lower than that of the ARIS-assisted transmission scenario because ARIS can directly amplify the incident signal. Meanwhile, under different maximum power amplification factors, ARIS only needs 202 and 235 unit cells to achieve the best performance, which is only about one-quarter to one-fifth of the unit cells needed by PRIS. With fewer unit cells, ARIS has a smaller surface size and is more suitable for space-constrained situations, such as indoor scenes. In addition, as the number of unit cells increases, the achievable rate of the ARIS-assisted transmission scenario first increases and then decreases after reaching the optimum point, which reflects the trade-off between the number of unit cells and the amplification power. ARIS can enhance the incident signal by increasing the number of unit cells and the amplification power. For a limited amplification power budget, increasing the number of unit cells can increase the achievable data rate. However, when the number of unit cells is greater than the optimal number of unit cells, the achievable data rate will decrease because more unit cells reduce the amplified power. This is very different from the common sense of PRIS that more unit cells always increase the data rate. So, the number of unit cells is an important variable that needs to be carefully considered if the RIS can enhance the incident signal. For a given ARIS power amplification budget, a higher maximum amplification factor $\beta _{\max }^2$ contributes to better performance. This is because a higher maximum amplification factor $\beta _{\max }^2$ has fewer amplification power constraints, and more power can be used to amplify the incident signal, thus achieving a larger SNR gain.

When $\beta _{\max }^2 =$ 40 dB, the number of active unit cells is small, and the achievable data rate of HRIS-assisted transmission scenario is greater than that of ARIS-assisted transmission scenario. This is because under a given RIS power budget, when the number of active unit cells is small, the amplification power in ARIS is not fully utilized, and the amplifier is in saturation. However, in HRIS, after the active unit cells amplify the incident signal with the maximum amplification factor, the remaining amplification power can be used to increase the passive unit cells so that the achievable data rate of the HRIS-assisted transmission scenario is larger than that of the ARIS-assisted transmission scenario. Meanwhile, as the number of active unit cells gradually increases, the remaining amplification power in HRIS gradually decreases so that the achievable data rate of HRIS-assisted transmission scenario gradually approaches that of ARIS-assisted transmission scenario. After reaching the optimum point, the achievable data rate of the HRIS-assisted transmission scenario and the ARIS-assisted transmission scenario overlap.
\begin{figure}[!t]
\centering
\includegraphics[width=3.2in]{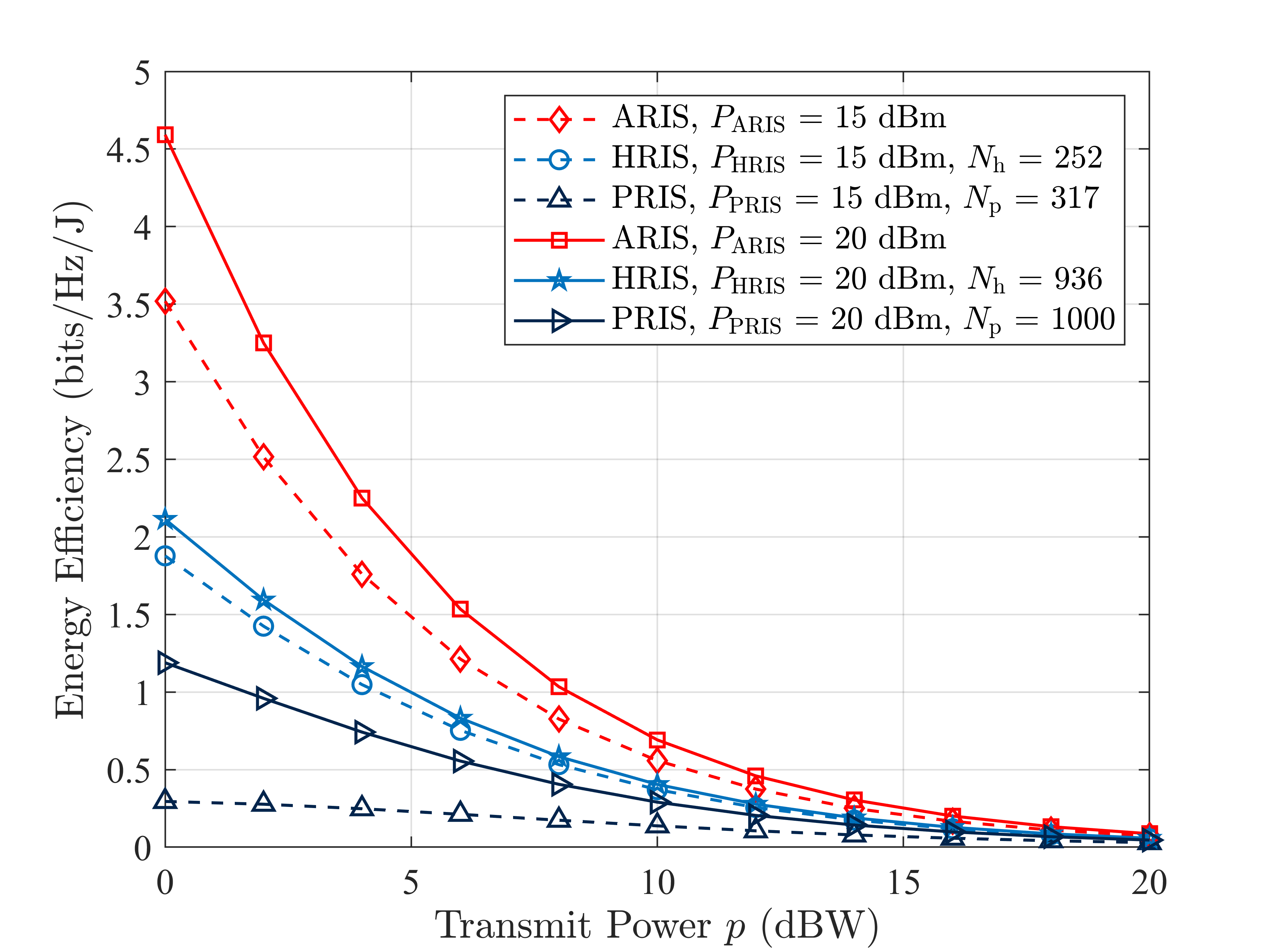}
\caption{Transmit power versus energy efficiency.}
\label{fig07}
\end{figure}

Next, this paper discusses the relationship between transmit power and energy efficiency, as shown in Fig.~\ref{fig07}. The simulation parameters are ${d_{{\text{st}}}} =$ 12 m, ${N_{\text{t}}} =$ 400, ${N_{{\text{ha}}}} =$ 20, ${P_{\text{s}}} = {P_{\text{d}}}=$ 20 dBm, ${P_{\text{e}}} =$ -10 dBm, ${P_{{\text{DC}}}} =$ -5 dBm, $v= $ 0.5, $\beta _{\max }^2 =$ 40 dB, ${P_{{\text{ARIS}}}} = {P_{{\text{HRIS}}}} = {P_{{\text{PRIS}}}} =$ 20 and 15 dBm with a bandwidth of $B =$ 10 MHz corresponding to a noise power of -94 dBm.

It can be seen from Fig.~\ref{fig07} that as $p$ increases, the energy efficiency of PRIS, ARIS, and HRIS-assisted transmission scenarios all decrease; the ARIS-assisted transmission scenario achieves the highest energy efficiency, and the PRIS-assisted transmission scenario obtains the lowest energy efficiency. This is because under the same transmit power $p$ and the same RIS power budget, the PRIS, ARIS, and HRIS-assisted transmission scenarios have the same total power. However, the achievable data rate of the ARIS-assisted transmission scenario is the largest, and the achievable data rate of the PRIS-assisted transmission scenario is the smallest. According to (56), the ARIS-assisted transmission scenario achieves the highest energy efficiency. When the power budget of the RIS is increased, the energy efficiency of the PRIS, ARIS, and HRIS-assisted transmission scenarios all increase. The reason is that when the power budget of the RIS is increased, the data rate increase of the PRIS, ARIS, and HRIS-assisted transmission scenarios is greater than the total power increase, thereby improving the energy efficiency.
\begin{figure}[!t]
\centering
\includegraphics[width=3.2in]{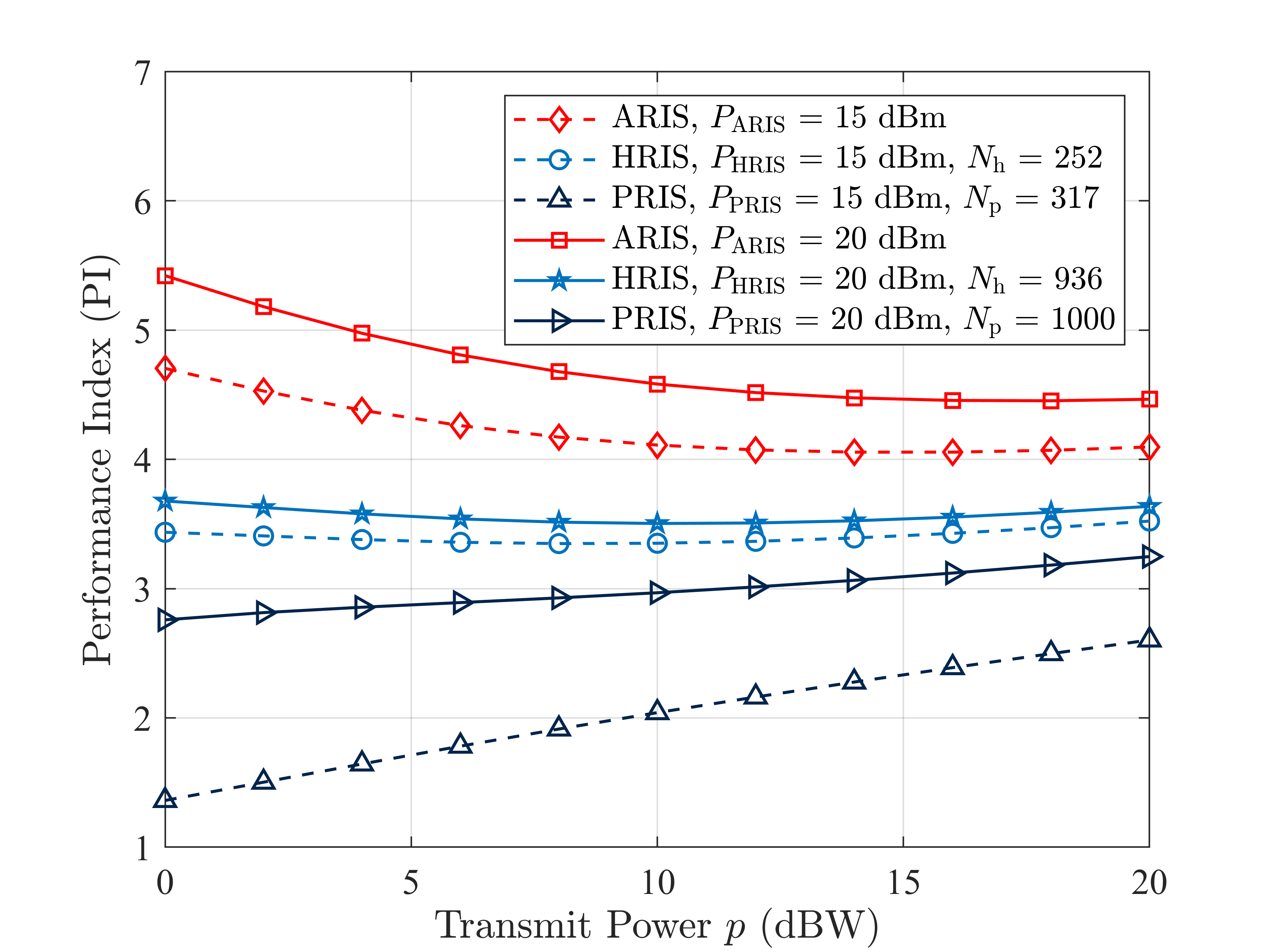}
\caption{PI versus transmit power, where $\lambda  = 0.5$.}
\label{fig08}
\end{figure}

Then, this paper investigates the relationship between PI and transmit power according to the formula for calculating the relationship between spectral efficiency and energy efficiency. The simulation parameters are the same as those in the above. When $\lambda  =$ 0.5, PI takes a compromise between energy efficiency and spectral efficiency, which is suitable for most scenarios.

It can be seen from Fig.~\ref{fig08} that the PI of the ARIS-assisted transmission scenario is the largest because the ARIS-assisted transmission scenario is larger than the PRIS and HRIS-assisted transmission scenarios regardless of the spectral efficiency or energy efficiency. As $p$ increases, the PI of the ARIS-assisted transmission scenario gradually decreases because of the upward trend of the achievable data rate with the increase of the power $p$ is smaller than that of the energy efficiency. The PI of the ARIS-assisted transmission scenario gradually decreases, indicating that the transmit power is a crucial variable when designing the achievable data rate and energy efficiency of the ARIS-assisted transmission scenario. Similarly, as $p$ increases, the PI of the HRIS-assisted transmission scenario first decreases and then increases, and there is a minimum value. This is because there is a minimum point in the compromise between the achievable data rate and energy efficiency of the HRIS-assisted transmission scenario. This reminds us to avoid this nadir when designing the HRIS-assisted transmission scenario. As $p$ increases, the PI of the PRIS-assisted transmission scenario gradually increases. This is mainly because the increase in the achievable data rate of the PRIS-assisted transmission scenario is more than the decrease in the energy efficiency. From the perspective of the RIS power budget, increasing the RIS power budget increases the PI of ARIS, HRIS, and PRIS-assisted transmission scenarios. The reason is that when increasing the RIS power budget, both the data rate and energy efficiency increase, and thus the PI increases.

\section{Conclusion}
This paper proposes a novel RIS-assisted mmWave indoor enhancement scheme where a transparent RIS is used for mmWave indoor signal enhancement. Three assisted transmission scenarios of PRIS, ARIS, and HRIS are used to transmit the signals to the UE, and HRIS is a novel RIS proposed in this paper to balance the hardware cost and channel gain. Specifically, given the RIS power budget, this paper considers maximizing the received SNR of the scheme to maximize the data rate. The closed-form maximum SNR is presented in the PRIS and ARIS-assisted transmission scenarios and that for given active unit cells is presented in HRIS-assisted transmission scenario. Then, this paper analyzes the performance of the proposed scheme under three assisted transmission scenarios. The results show that under a specific RIS power budget, the ARIS-assisted transmission scenario achieves the highest achievable data rate and energy efficiency. Meanwhile, the number of unit cells required in ARIS is about a quarter of that of PRIS or HRIS, which dramatically reduces the volume of the metasurface and is more suitable for space-constrained situations, such as indoor scenes. It should be noted that the ARIS-assisted transmission scenario balances the number of unit cells and the amplification power, which is different from PRIS. So, the number of unit cells needs to be carefully designed when the ARIS is used for a connection. Furthermore, under the same RIS power budget, the achievable data rate and energy efficiency under the HRIS-assisted transmission scenario are greater than those under the PRIS-assisted transmission scenario. These results indicate that HRIS can better balances performance and cost.


\newpage



\vspace{11pt}


\vfill

\end{document}